\documentclass[
prl
,twocolumn
,nofootinbib
,floatfix
,superscriptaddress
]{revtex4-2} 

\usepackage{notes2bib}
\usepackage{natbib}
\usepackage{gensymb}
\usepackage{amsthm}
\usepackage{amsmath}
\usepackage{color}
\usepackage{bm}
\usepackage{amsfonts}
\usepackage{dsfont}
\usepackage{graphicx}
\usepackage{float}
\usepackage{subfigure}
\usepackage{verbatim}
\usepackage{amsfonts}
\usepackage{bbold}
\usepackage{dcolumn}
\usepackage{bm}
\usepackage[dvipsnames]{xcolor}
\usepackage{listings}
\definecolor{blueprl}{RGB}{46,48,146}
\usepackage[pdftex,colorlinks=true,urlcolor=blueprl,citecolor=blueprl,linkcolor=blueprl]{hyperref}
\usepackage{mathrsfs}
\usepackage{mathtools}
\usepackage{times}
\usepackage{color}
\usepackage[dvipsnames]{xcolor}

\usepackage{braket}

\usepackage{soul}

\newcommand{\nonvt}{\vphantom{\frac{1}{\sqrt{2}}} \nonumber} 
\newcommand{\mrm}[1]{\mathrm{#1}}

\usepackage{mathtools}
\usepackage{dsfont}
\usepackage[mathscr]{euscript}
\usepackage{amsmath}
\usepackage{amssymb}
\usepackage{graphicx}
\usepackage{dcolumn}
\usepackage{bm}
\usepackage{epsfig}
\usepackage{latexsym,mathrsfs}
\usepackage{graphicx}
\usepackage{color}
\usepackage{hyperref}
\usepackage{float}
\usepackage{diagbox}
  \usepackage{cancel}
  \usepackage[normalem]{ulem}

\definecolor{vividviolet}{rgb}{0.62, 0.0, 1.0}
\definecolor{amaranth}{rgb}{0.9, 0.17, 0.31}
\definecolor{palatinateblue}{rgb}{0.15, 0.23, 0.89}
\definecolor{brightpink}{rgb}{1.0, 0.0, 0.5}
\definecolor{cornflowerblue}{rgb}{0.39, 0.58, 0.93}
\definecolor{deepcarminepink}{rgb}{0.94, 0.19, 0.22}
\definecolor{radicalred}{rgb}{1.0, 0.21, 0.37}
\definecolor{blueblue}{RGB}{21,47,181}
\definecolor{greengreen}{RGB}{65,166,16}

\usepackage{xtab,afterpage}
\usepackage{hyperref}
\usepackage{setspace,calc}

\usepackage[pdftex,colorlinks=true,urlcolor=blueprl,citecolor=blueprl,linkcolor=blueprl]{hyperref}

\newcommand{\be}{\begin{equation}}
\newcommand{\ee}{\end{equation}}
\newcommand{\bs}{\begin{split}} 
\newcommand{\bea}{\begin{eqnarray}}
\newcommand{\eea}{\end{eqnarray}}

\newcommand{\vt}{\vphantom{\frac{1}{2}}}

\newcommand{\non}{\nonumber} 
 
\newcommand{\p}{\partial} 
\newcommand{\D}{\mathrm{d}}
\newsavebox{\myhbar}

\newcommand{\wv}[3]{{}_{#1}\langle {#2}_w \rangle_{#3}}
\newcommand{\dd}{\dagger}
\newcommand{\mtx}{\mathbf{X}}

\allowdisplaybreaks

\begin{document}

\newcommand{\CQCT}{\affiliation{Centre for Quantum Computation \& Communication Technology, School of Mathematics \& Physics, The University of Queensland, St.~Lucia, Queensland, 4072, Australia}}
\newcommand{\FUB}{\affiliation{Dahlem Center for Complex Quantum Systems, Freie Universit\"at Berlin, 14195 Berlin, Germany}}
\newcommand{\STE}
{\affiliation{Department of Physics, Stevens Institute of Technology, Castle Point Terrace, Hoboken, New Jersey 07030, U.S.A.}}

\title{Measurement-based Lorentz-covariant Bohmian trajectories of interacting photons}
\author{Joshua Foo}
\email{joshua.foo@uqconnect.edu.au}
\CQCT
\STE 

\author{Austin P. Lund}
\CQCT
\FUB

\author{Timothy C. Ralph}
\email{ralph@physics.uq.edu.au}
\CQCT

\begin{abstract}
In a recent article [Foo et.\ al., \textit{Nature Comms.} \textbf{13}, 2 (2022)], we devised a method of constructing the Lorentz-covariant Bohmian trajectories of single photons via weak measurements of the photon's momentum and energy. However, whether such a framework can consistently describe multiparticle interactions remains to be seen. Here, we present a nontrivial generalisation of our framework to describe the relativistic Bohmian trajectories of two interacting photons exhibiting nonclassical interference due to their indistiguishability. We begin by deriving the average velocity fields of the indistinguishable photons using a conditional weak measurement protocol, with detectors that are agnostic to the identity of the respective photons. We demonstrate a direct correspondence between the operationally-derived trajectories with those obtained using a position- and time-symmetrised multiparticle Klein-Gordon wavefunction, whose dynamics are manifestly Lorentz-covariant. We propose a spacetime metric that depends nonlocally on the positions of both particles as a curvature based interpretation of the resulting trajectories. Contrary to prior expectations, our results demonstrate a consistent trajectory-based interpretation of relativistic multiparticle interactions in quantum theory.
\end{abstract}

\date{\today}

\maketitle

Bohmian mechanics is a deterministic, nonlocal interpretation of quantum mechanics that posits the existence of real particle trajectories.\ Unlike so-called \textit{epistemological} approaches like the Copenhagen interpretation, which are primarily interested in what may be known about a given quantum system, Bohmian mechanics is an \textit{ontological} approach to quantum theory, since it is primarily interested in the nature or reality of a quantum system \cite{bohm2006undivided,bohmPhysRev.85.166}. As is required of any interpretation, it recovers many of the standard results of nonrelativistic quantum mechanics such as the Born rule, where the square of the wavefunction,
\begin{align}
    \rho(t,x) &= | \psi(t,\mathbf{x}) |^2,
\end{align}
is interpreted as the probability density of finding a particle at the spacetime point $(t,\mathbf{x})$ \cite{sanz2012trajectory}. In Bohmian mechanics, $\rho(t,\mathbf{x})$ is interpreted as the density of particle trajectories given some initial distribution, which is conserved for all future times as required by the continuity equation, 
\begin{align}
    \frac{\p \rho(t,\mathbf{x})}{\p t} &= - \nabla \cdot j (t, \mathbf{x}) 
\end{align}
where $j(t,\mathbf{x})$ is the conserved probability current.

Interest in Bohmian mechanics was revived following the paper by Wiseman \cite{Wiseman_2007}, who showed that the standard Bohmian velocity field, $v(t,\mathbf{x}) = j(t,\mathbf{x})/\rho(t,\mathbf{x})$, could be determined operationally via weak measurements.\ A weak measurement is a method of performing unbiased measurements on quantum systems with minimal disturbance due to the measurement process itself \cite{aharonovPhysRevLett.60.1351,dresselRevModPhys.86.307,lundeenPhysRevLett.108.070402,tamir2013introduction}. Specifically, a weak measurement of the observable $\hat{a}$ only weakly perturbs the system of interest, but concomitantly carries a large amount of measurement uncertainty. Performing repeated weak measurements on an ensemble scales this uncertainty as $1/\sqrt{N}$ (where $N$ is the number of measurements) allowing one to estimate the average value $\langle \hat{a} \rangle$ with arbitrarily high precision \cite{dresselRevModPhys.86.307}. 

A ``weak value'' extends this notion of weak measurement by introducing a subsequent strong measurement and performing post-selection on the ensemble based on the outcome of this strong measurement. 
Formally, the weak value of $\hat{a}$, denoted $\prescript{}{\langle \phi|}{\langle \hat{a}_w} \rangle_{|\psi\rangle}$, is the mean value of $\hat{a}$ obtained from many weak measurements on an ensemble of particles each prepared in the state $|\psi\rangle$, postselecting only those particles where a later strong measurement reveals the system to be in the state $|\phi\rangle$. This results in the definition of a weak value given by the formula \cite{aharonovPhysRevLett.60.1351}:
\begin{align}\label{eqwv4}
    \wv{\bra{\phi}}{\hat{a}}{\ket{\psi(t)}} =
    \mathrm{Re} \frac{\langle \phi | \hat{a} | \psi(t) \rangle }{\braket{\phi|\psi(t)}}.
\end{align}
From this starting point, Wiseman identified the Bohmian velocity field of single nonrelativistic particles with the weak value of the particle's momentum divided by its mass:
\begin{align}\label{eq4}
    v(t,\mathbf{x}) &= \frac{\wv{\bra{x}}{\hat{p}}{\ket{\psi(t)}} }{m} ,
\end{align}
where the postselection occurs on particles found in the position eigenstate $| x \rangle$. Similar proposals have been made by Leavens \cite{Leavens2005} and Hiley \cite{hiley_2012}\nocite{Matzkin_2015}. The velocity equation Eq.\ (\ref{eq4}) is strictly an \textit{average} obtained over post-selected outcomes on an ensemble \cite{FANKHAUSER202116,Bliokh_2013}. An additional assumption of determinism links Eq.\ (\ref{eq4}) with the framework of Bohmian mechanics--the average trajectories of the particles are understood to correspond to actual (realistic) trajectories in spacetime. A significant achievement of Bohmian mechanics and Wiseman's weak value framework has been the experimental inference of the trajectories of both single and entangled photons in the nonrelativistic limit \cite{kocsis2011observing,mahler2016experimental,Xiao:17,bravermanPhysRevLett.110.060406}. By nonrelativistic, we mean that the longitudinal degrees of freedom of the measured photons are non-dynamic and the transverse velocities are slow, allowing for the paraxial approximation $E(k) \simeq k_z + k^2/(2k_z)$ to be invoked, where $E(k)$ is the photon energy and $k$, $k_z$ are its momenta in the transverse and longitudinal directions respectively.

In a recent article, we reformulated Wiseman's nonrelativistic framework to incorporate relativistic effects for single photons \cite{Foo_2022Rel}.\ In particular, we constructed the relativistic velocity field of a single particle in terms of weak values of its momentum and energy (and specialising to a (1+1)-dimensional setup):
\begin{align}\label{eq3}
    v(t,x) &= \frac{\wv{\bra{x}}{(\hat{p}_x)}{\ket{\psi(t)}} }{\wv{\bra{x}}{\hat{H}}{\ket{\psi(t)}} } .
\end{align}
We showed that this weak value velocity field can be re-expressed in terms of the components of the scalar Klein-Gordon conserved current (when making the simplification to a spin-0 dynamical theory), giving a Lorentz covariant quantity obeying relativistic velocity addition (the measurement framework does not depend on the specific form of the relativistic Hamiltonian or the initial states considered). In light of the historical difficulties associated with the scalar Klein-Gordon equation (and more generally, particle interpretations of relativistic wave equations) \cite{Ghose_2001,BOHM1987321,Struyve:2004xd,berndlPhysRevA.53.2062,horton2002broglie,2000horton,nikolic2004,nikolic2005,nikolic2005_2,Berry_2012,Alkhateeb2022,durrdoi:10.1098/rspa.2013.0699}, the achievement of Eq.\ (\ref{eq3}) is the grounding of relativistic single particle Bohmian trajectories in operational measurements, which can be in-principle verified experimentally. 

\begin{figure}[h]
    \centering
    \includegraphics[width=0.6\linewidth]{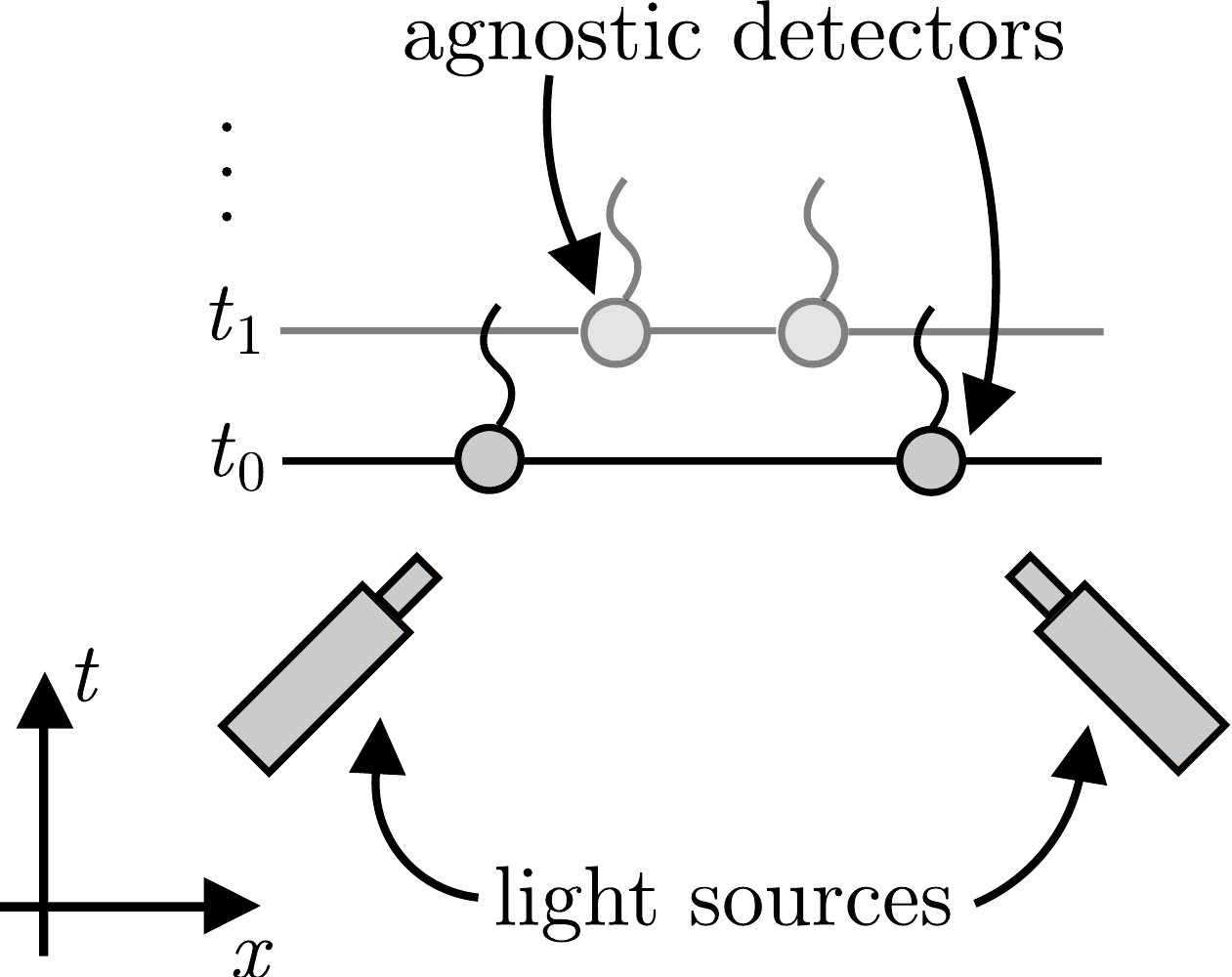}
    \caption{Schematic diagram of the system of interest. Two single-photon light sources are directed towards each other, with the detectors, operating on an equal timeslice, performing weak measurements of the detected photons.}
    \label{fig:schematic}
\end{figure}

Nevertheless, it remains to be seen whether such a weak value formulation is valid beyond the single-particle limit. Motivated by this question, we provide a nontrivial extension of our weak value framework and construct the operationally-determined Bohmian trajectories of two interacting photons. We ground the trajectories in weak measurements of each photon's momentum and energy (what we refer to as a ``measurement-based approach'', based in the usual field-theoretic notions of state preparation, time-evolution, and measurement), before drawing a direct connection with a manifestly Lorentz-covariant multiparticle Klein-Gordon theory for scalar particles \cite{pladevall2019applied,mandel1995optical}.\ Beginning with our measurement-based approach, we consider an initially separable state of two photons directed towards each other with oppositely-directed wavevectors. That is, unlike the paraxial approximation, the relativistic limit $E(k) = \sqrt{k^2 + k_z^2} \simeq | k|$ (where $k \equiv k_x$ is the momentum in the transverse or $x$-direction) is taken. We consider two position detectors that are agnostic to the directionality of the wavevectors of any incident photons \cite{contopoulosparticles6010005}, with both of them operating along a single timeslice (i.e.\ spacelike hypersurface, see Fig.\ \ref{fig:schematic}). 
Treating the detectors this way erases which path information, provided the photons are indistinguishable. Next, we consider an equivalent scenario using the multiparticle Klein-Gordon equation, where the initial state is symmetrised in time and position. Our motivation for this is to ground our operationally derived trajectories in a ``formal'' dynamical theory. Our utilisation of the scalar theory captures the essential Lorentz relativistic properties of photons, with the understanding that it is a simplification of a full electrodynamical theory \cite{bialynicki1994wave,BIALYNICKIBIRULA1996245}. The deterministic velocity fields of the two photons are constructed by computing the time and space components of the conserved current vector, these components satisfying a continuity equation for the respective photons. Ghose et.\ al.\ have considered a similar scenario (without drawing a connection with weak values) using the modified dynamics of the Kemmer equation \cite{Ghose_2001}, although in their formulation they explicitly consider a paraxial-type approximation for the trajectories in the Fraunhofer limit, and the resulting velocity field is timelike everywhere (unlike our construction using the scalar Klein-Gordon equation in the fully relativistic limit, which admits spacelike tangent vectors) \cite{STRUYVE200484}.

Our main result is that the photon velocities and trajectories obtained using either the weak value or Klein-Gordon approach are identical. In other words, we have grounded the relativistic Bohmian trajectories of multiple interacting photons in an operational model based in weak measurements, which in-principle, allows such trajectories to be detected in experiment. Contrary to prior expectations, our results demonstrate that there is a consistent way of understanding multiparticle interactions in harmony with a trajectory-based ontology and alongside the tenets of special relativity. Of course, to be consistent with Bell inequalities, nonlocal parameter dependence is expected to emerge. How this becomes manifest will be discussed in the Methods section.

Our article is organised as follows.\ We first present a derivation of the two-photon velocity fields using our measurement-based approach.\ We show that the same result can be obtained by beginning with a multiparticle scalar Klein-Gordon theory. We plot the relativistic Bohmian-type trajectories of the two photons and discuss the new features of these trajectories, before concluding with some final remarks. Throughout this article we adopt natural units, i.e.\ $\hslash = c = 8G = 1$. 

\section{Results}
\subsection{Measurement-Based Approach}

Let us first derive the velocity fields and trajectories for the system of interest from our measurement-based approach. As mentioned, we consider an initially separable state of the two photons:
\begin{align}\label{eq5.12}
    | \psi(t) \rangle &= \hat{a}_{f_1}^{\dd} \hat{a}_{f_2}^{\dd} | 0 \rangle | 0 \rangle = | f(k_1) \rangle | f(k_2) \rangle ,
\end{align}
where 
\begin{align}\label{eq2}
    \hat{a}_{f_i}^{\dd} | 0 \rangle &= \int\D k_i \: e^{-iE (k_i)t} f(k_i) \hat{a}_{k_i}^{\dd} | 0 \rangle .
\end{align}
Here $\hat{a}_{k_i}^\dd$ is the creation operator that creates a photon in the plane wave mode $k_i$, i.e.\ $\hat{a}_{k_i}^\dd | 0 \rangle \equiv | k_i \rangle$. $E(k_i) = \sqrt{k^2 + k_z^2} \simeq |k|$ is the energy of the photon in mode $k_i$, having made the simplifying assumption $k_i\gg k_z$. The operator $\hat{a}_{f_i}^\dd$ creates a photon in the mode $k_i$ with a distribution of frequencies given by the Gaussian wavepacket
\begin{align}
    f_i(k_i) &= \mathcal{N} \exp \left[ - \frac{(k-k_0)^2}{4\sigma^2} \right],
\end{align}
where $\mathcal{N}$ is a normalisation constant, $k_0$ is the centre frequency and $\sigma$ the bandwidth. Such a plane wave description is a good description of a transverse Gaussian mode close to its beam waist. If $|k_0|$ and $\sigma$ are the same for the two photons, they will be indistinguishable apart from their propagation direction. If in addition, as discussed, the detectors are non-directional then interference effects between the photons can appear. We then construct the photon velocity fields via weak values of observables measured by the detectors. The strong measurement enacted by the detectors only reveals information about the photon positions.\ We are interested in the case where each detector registers a single detection event, and focus on the case where the photons have initially oppositely directed wavevectors (i.e.\ one left- and one right-moving photon). 

In view of this, we propose the following definitions for the velocity fields of the two photons, as constructed via weak measurements of the photon momenta and energy locally at detector $A$ and detector $B$:
\begin{align}\label{eq5.15}
    v_1(x_1,x_2,t) &= \frac{\prescript{}{\langle \bar{x} | }{\langle \hat{\mathbf{k}}_{Aw} \rangle}_{|\psi(t)\rangle}}{\prescript{}{\langle \bar{x} |}{\langle \hat{\mathbf{H}}_{Aw} \rangle }_{|\psi(t)\rangle}} , 
    \\
    \label{eq5.16}
    v_2(x_1,x_2,t) &= \frac{\prescript{}{\langle \bar{x} | }{\langle \hat{\mathbf{k}}_{Bw} \rangle}_{|\psi(t)\rangle}}{\prescript{}{\langle \bar{x} |}{\langle \hat{\mathbf{H}}_{Bw} \rangle }_{|\psi(t)\rangle}} .
\end{align}
Here, $| \bar{x} \rangle$ describes the strong measurement made by two detectors $A$ and $B$ on the positions of both photons, while $\hat{\mathbf{k}}_D$, $\hat{\mathbf{H}}_D$ are the momentum and Hamiltonian operators associated with detectors $A$ and $B$. From a field-theoretic standpoint, it is natural to consider the two detectors as operating on a single timeslice $t$, enacting weak measurements of the detected momentum and energy of the incident photons, and postselecting the outcomes from an ensemble of trials based on the strong measurement of the photon positions in the state $|\bar{x}\rangle$. Over repeated runs of the experiment, the detectors incrementally evolve through successive timesteps, which allows them to determine, for a given position and time of arrival of the two photons, the weak values of momentum and energy needed to construct the velocity fields shown in Eq.\ (\ref{eq5.15}) and (\ref{eq5.16}) (see Fig.\ \ref{fig:1}). In post-processing, the measurement results are used to interpolate the trajectories traversed by each photon given some initial conditions. We emphasise here that detector $A$ does not detect ``photon 1'' exclusively, or vice versa. Within the Bohmian interpretation, the particle trajectories are deterministic and do not cross. If we place our detectors in the region prior to any interaction between the photons, with (say) detector $A$ on the left and detector $B$ on the right, then we can identity $v_1(t,x_1,x_2)$ with the ``velocity detected by detector $A$'' and vice versa. If detector $B$ was on the left, we would make the opposite identification. However, in the interference region where the photons begin interacting (shown for example in Figs.\ \ref{fig:1} and \ref{fig:2bohm}), the detectors cannot identify which photon they measured. In order to tell the photons apart, one needs to interpolate the trajectories from the measurement outcomes of both detectors--performed on incrementally successive timeslices--at the end of the proposed experiment. 

We also remark that our detector model does not constrain us to perform measurements on an equal timeslice; one can in-principle configure the detectors so that they are timelike separated. However, the resulting outcomes, and consequently the trajectories, will be completely different in such a scenario. For example, if detector $A$ is operative in a finite spacetime region in the past lightcone of detector $B$, then the backaction of detector $A$'s conditional measurement will inevitably affect the outcome of detector $B$'s measurement. Thus, with the restriction that both detectors are compactly supported in spacelike regions of spacetime, then the choice of measurements performed on an equal timeslice is the conceptually simplest one (noting also that a boost can transform any scenario of spacelike separated detectors operating on \textit{unequal} timeslices into one in which they operate on an equal timeslice).

Let us return to Eq.\ (\ref{eq5.15}) and (\ref{eq5.16}). The weak values are explicitly given by 
\begin{align}\label{eq5.17}
    \prescript{}{\langle \bar{x} | }{\langle \hat{\mathbf{k}}_{Dw} \rangle}_{|\psi(t)\rangle} &= \mathrm{Re} \frac{\langle \bar{x} | \hat{\mathbf{k}}_D | \psi(t)\rangle}{\langle \bar{x} | \psi(t) \rangle} , 
    \\
    \label{eq5.18}
    \prescript{}{\langle \bar{x} | }{\langle \hat{\mathbf{H}}_{Dw} \rangle}_{|\psi(t)\rangle} &= \mathrm{Re} \frac{\langle \bar{x} | \hat{\mathbf{H}}_D | \psi(t) \rangle}{\langle \bar{x} | \psi(t) \rangle} .
\end{align}
Here, $| \bar{x} \rangle$ describes the joint state of the two detectors, each sensitive to a single photon in either the $k_1$ or $k_2$ mode but insensitive to the identity of the detected photon (we emphasise that this is the same detector model that was tacitly assumed in the single-photon case).\ This can be obtained by considering an initially separable state of the detectors, $| \bar{x} \rangle = | x_A \rangle | x_B \rangle$, where the states $|x_{A,B} \rangle$ are individually constructed as superpositions of $x$-eigenstates with support on $k<0$ and $k>0$ respectively (i.e.\ to account for detections from the left- and right-moving directions). We note here that $| x_{A,B} \rangle$ can be understood as a smeared position state of the detector in the narrowband limit. Moreover, these states model measurements performed in a particular reference frame (originally, the lab frame); if one wants to consider trajectories in a boosted reference frame, one can perform a Lorentz transformation of the measured velocities to obtain those in the new frame. Conversely, if we move into a different frame and apply our operational definition, Eq.\ (\ref{eq5.15}) and (\ref{eq5.16}), we can relate the velocities we obtain in this frame to those that would be obtained in the original frame through a standard Lorentz transformation.

Since there are two photons, upon taking the direct product $|x_A \rangle |x_B\rangle$ we postselect on the detector coincidences such that detector $A$ measures the photon in mode $k_1$ while detector $B$ measures the photon in mode $k_2$, or vice versa. After renormalising this conditional state, one obtains 
\begin{align}
    | \bar{x} \rangle
    \label{eq5.19}
    &= \frac{1}{\sqrt{2}} \left( | x_{1}, k_{1_A} \rangle | x_{2}, k_{2_B} \rangle + | x_1 , k_{2_A} \rangle | x_2, k_{1_B} \rangle \right) ,    
\end{align}
where we have adopted the nomenclature
\begin{align}
    | x_i, k_{j_D} \rangle &= \int\D k_{j_D} e^{-ik_{j_D}x_i} |k_{j_D} \rangle.  
\end{align}
Here $i,j = 1,2$ labels the photon of interest, $D = A,B$ labels the respective detectors, and $x_i$ is the position coordinate associated with photon $i$. Equation (\ref{eq5.19}) describes the fact that the photons are indistinguishable. Since either detector is sensitive to a single detection event from either direction, then detector $A$ will detect a photon in mode $k_1$ while detector $B$ detects a photon in mode $k_2$, \textit{or} detector $A$ detects a photon in mode $k_2$ while detector $B$ detects a photon in mode $k_1$. A similar symmetrisation procedure also applies to the momentum and Hamiltonian operators in Eq.\ (\ref{eq5.17}) and (\ref{eq5.18}). That is, the detectors measure ``a momentum'' and ``an energy'' in their respective localised spacetime regions, however they remain agnostic to the identity of the detected photon. 

This motivates us to consider operators associated with the observables of each detector, given by 
\begin{align}
    \hat{\mathbf{k}}_A &= \hat{k}_A | k_{1_A} \rangle\langle k_{1_A} | \otimes \hat{I}_B |k_{2_B} \rangle\langle k_{2_B} | 
    \non 
    \\
    & \qquad + \hat{k}_{A} | k_{2_A} \rangle\langle k_{2_A} | \otimes \hat{I}_B | k_{1_B} \rangle\langle k_{1_B} |  \vt ,
    \\
    \hat{\mathbf{k}}_B &= \hat{k}_B | k_{1_B} \rangle\langle k_{1_B} | \otimes \hat{I}_A | k_{2_A} \rangle\langle k_{2_A} | 
    \non 
    \\
    & \qquad + \hat{k}_B | k_{2_B} \rangle\langle k_{2_B} | \otimes \hat{I}_A | k_{1_A} \rangle\langle k_{1_A} | ,
    \vt 
    \\
    \label{eq17}
    \hat{\mathbf{H}}_A &= \hat{H}_A | k_{1_A} \rangle\langle k_{1_A} | \otimes\hat{I}_B | k_{2_B} \rangle\langle k_{2_B} | 
    \non 
    \\
    & \qquad + \hat{H}_A | k_{2_A} \rangle\langle k_{2_A} | \otimes \hat{I}_B | k_{1_B} \rangle\langle k_{1_B} | ,
    \vt 
    \\
    \label{eq18}
    \hat{\mathbf{H}}_B &= \hat{H}_B | k_{1_B} \rangle\langle k_{1_B} | \otimes\hat{I}_A | k_{2_A} \rangle\langle k_{2_A} | 
    \non 
    \\
    & \qquad 
    + \hat{H}_B | k_{2_B} \rangle\langle k_{2_B} | \otimes \hat{I}_A | k_{1_A} \rangle\langle k_{1_A} | . 
    \vt 
\end{align}
Again, these operators may be understood as being formed from weak value measurements at detector $A$, given a coincident strong value measurement at detector $B$, and vice versa. $\hat{k}_D$, $\hat{H}_D$, and $\hat{I}_D$ are the momentum, Hamiltonian, and identity operators with support on the Hilbert space of detector $D = A, B$. To understand this prescription, let us take $\hat{\mathbf{k}}_A$ for example. This operator is associated with the momentum detected by detector $A$, and thus has no support on the Hilbert space associated with detector $B$.\ Taking an ideal case where the detector is in the state $| k_{1_A} \rangle | k_{2_B} \rangle$, the measurement outcome is 
\begin{align}
    \langle k_{1_A} | \langle k_{2_B} | \hat{\mathbf{k}}_A | k_{2_B} \rangle | k_{1_A} \rangle &= k_{1_A} ,
    \vt 
    \vphantom{\sqrt{\frac{2}{\pi}}}
\end{align}
while for the state $| k_{2_A} \rangle | k_{1_B} \rangle$, 
\begin{align}
    \langle k_{2_A} | \langle k_{1_B} | \hat{\mathbf{k}}_A | k_{1_B} \rangle | k_{2_A} \rangle &= k_{2_A}  . 
    \vt 
    \vphantom{\sqrt{\frac{2}{\pi}}}
\end{align}
With these ingredients, it is now possible to compute the different components of the weak value velocity fields explicitly. As discussed, let us consider the ``head-on'' case where $k\gg k_z$ leading to the dispersion relation $E(k) \simeq | k |$. In this limit, and using Eq.\ (\ref{eq5.12}) and (\ref{eq5.19}), the position-space wavefunction $\psi_\mathrm{M} \equiv \psi_\mathrm{M}(t,x_1,x_2) = \langle \bar{x} | \psi(t) \rangle$ is given by:
\begin{align}
    & \psi_\mathrm{M} \equiv \psi_\mathrm{M}(t,x_1,x_2) = \langle \bar{x} | \psi (t) \rangle \non
    \\
    &\:\:\: = \frac{1}{\sqrt{2}} ( \psi_1(t,x_1) \psi_2(t,x_2) 
    + \psi_1(t,x_2) \psi_2(t,x_1) ), 
    \vphantom{\sqrt{\frac{2}{\pi}}}
\end{align}
where
\begin{align}
    \psi_1(t,x_i) &= \int\D k \: e^{-ik (t- x_i)} f(k;k_0) ,
    \\
    \psi_2(t,x_i) &= \int\D k \: e^{-ik(t+x_i)} f(k;k_0) . 
\end{align}
Here, $\psi_\mathrm{M}(t,x_1,x_2) = \psi_\mathrm{M}(t,x_2,x_1)$ is symmetric under an exchange of the photon positions. Meanwhile, the numerators for the respective weak value expressions are 
\begin{align}\label{eq5.31}
    & \langle \bar{x} | \hat{\mathbf{k}}_A | \psi(t) \rangle 
    \non 
    \\
    & \:\:\: 
    = \frac{1}{\sqrt{2}} \left( \psi_{1k} (t,x_1) \psi_2(t,x_2)  - \psi_1(t,x_2) \psi_{2k}(t,x_1) \right) ,
    \\
    & \langle \bar{x} | \hat{\mathbf{k}}_B | \psi(t) \rangle 
    \non 
    \\
    & \:\:\: = \frac{1}{\sqrt{2}} \left( \psi_{1k}(t,x_2) \psi_2(t,x_1) - \psi_1(t,x_1) \psi_{2k}(t,x_2) \right) ,
\end{align}
and the denominators are
\begin{align} 
    &\langle \bar{x} | \hat{\mathbf{H}}_A | \psi(t) \rangle 
    \non 
    \\
    & \:\:\: = \frac{1}{\sqrt{2}} \left( \psi_{1k}(t,x_1) \psi_2(t,x_2)  + \psi_1(t,x_2) \psi_{2k}(t,x_1) \right) ,
    \\
    \label{eq5.34}
    & \langle \bar{x} | \hat{\mathbf{H}}_B | \psi(t) \rangle 
    \non 
    \\
    & \:\:\: = \frac{1}{\sqrt{2}} \left( \psi_1(t,x_1) \psi_{2k}(t,x_2)  + \psi_{1k}(t,x_2) \psi_2(t,x_1) \right), 
\end{align}
having additionally defined
\begin{align}
    \psi_{1k}(t,x_i) &= \int\D k \: e^{-ik(t-x_i)}k f(k;k_0) ,
    \\
    \psi_{2k}(t,x_i) &= \int\D k \: e^{-ik(t+x_i)}k f(-k;k_0) .
\end{align}
These can be straightforwardly integrated to yield analytic expressions for the detector weak values, and hence the velocity fields of the two photons.  

\subsection{Multiparticle Klein-Gordon Approach}
We now wish to show that the measurement-based approach yields trajectories that are formally equivalent to those obtained using {a particular interpretation of relativistic quantum mechanics for multiple particles. The purpose of this is to confirm that the measurement-based trajectories satisfy a continuity constraint imposed by a manifestly Lorentz covariant dynamical theory, in analogy with our single-particle results. 

For simplicity, we adopt the multitime Klein-Gordon theory for multiple scalar particles presented in \cite{pladevall2019applied}. First, consider the position-space representations of a right- and left-moving single-particle Klein-Gordon wavefunction, for a Gaussian wavepacket of momenta $f(k;k_0)$, 
\begin{align}
    \psi_1(\mtx_j) &= \int\D k \: e^{-ik(t_j - x_j)} f(k;k_0) ,
    \\
    \psi_2(\mtx_j) &= \int\D k \: e^{-ik(t_j + x_j)} f(k;k_0) ,
\end{align}
where $\psi_1$, $\psi_2$ are associated with the respective photons. Notice in particular the introduction of a multitime identification of each photon's spacetime location, $\mtx_j = (t_j, x_j)$. Such an identification encapsulates the manifest Lorentz covariance of the relativistic dynamics, meaning that Lorentz transformations are enacted on each pair of the spacetime variables. For indistinguishable photons, one needs to symmetrise the wavefunction in position and time, yielding the wavefunction $\psi_\mathrm{KG}\equiv  \psi_\mathrm{KG}(\mtx_1,\mtx_2)$ of the two photons:
\begin{align}\label{eq5.54}
    &\psi_\mathrm{KG} \equiv \psi_\mathrm{KG}(\mtx_1, \mtx_2) 
    \non 
    \\
    & \:\:\: = \frac{1}{\sqrt{2}} \left( \psi_1(\mtx_1) \psi_2(\mtx_2) + \psi_1(\mtx_2) \psi_2(\mtx_1) \right) . 
\end{align}
The wavefunction Eq.\ (\ref{eq5.54}) satisfies two Klein-Gordon equations:
\begin{align}
    \p_{1\mu} \p^\mu_1 \psi(\mtx_1 , \mtx_2 ) &= 0 
    \vt ,
    \\
    \p_{2\mu} \p^\mu_2 \psi(\mtx_1, \mtx_2 ) &= 0 
    \vt ,
\end{align}
one for each pair of spacetime variables $\mtx_1$, $\mtx_2$. One thus obtains two conserved currents, 
\begin{align}
    j^\mu_1 &= 2 \mathrm{Im} \: \psi^\star( \mtx_1, \mtx_2 ) \p^\mu_1 \psi( \mtx_1, \mtx_2 ) , 
    \vt 
    \\
    j^\mu_2 &= 2 \mathrm{Im} \: \psi^\star(\mtx_1, \mtx_2) \p^\mu_2 \psi(\mtx_1, \mtx_2 ) ,
    \vt 
\end{align}
which individually satisfy a continuity equation, 
\begin{align}\label{eq5.43}
    \p_{\mu 1} j^\mu_1 &= 0 ,
    \vt 
    \\
    \label{eq5.44}
    \p_{\mu 2} j^\mu_2 &= 0 
    \vt .
\end{align}
To obtain the velocity field of each particle, one uses the Bohmian-type formulae (as motivated by our single-particle derivation), 
\begin{align}
\label{eq31}
    v_1(\mtx_1, \mtx_2) &= \frac{j_1(\mtx_1, \mtx_2)}{\rho_1(\mtx_1, \mtx_2)} , 
    \\
\label{eq32}
    v_2( \mtx_1, \mtx_2) &= \frac{j_2(\mtx_1, \mtx_2)}{\rho_2(\mtx_1, \mtx_2) }, 
\end{align}
where each velocity equation is formed from the components of the Klein-Gordon conserved current vector, $\mathbf{j}_i \equiv (j^0_i, j^\mu_i) = ( \rho_i , j_i ) $ for particle $i = 1, 2$. The satisfaction of the continuity equations Eq.\ (\ref{eq5.43}) and (\ref{eq5.44}) guarantees that the trajectory densities of the two photons are related to the probability density in the correct way (we show this graphically in ``Relativistic Two-Photon Trajectories'', and prove this in ``Methods'').  
To obtain a physical scenario yielding an equivalence with our measurement-based approach, we compute the velocities on a single timeslice:
\begin{align}\label{eq21}
    v_{1,t} &\equiv v_{1,t} (t,x_1,x_2) = v_1 (\mtx_1, \mtx_2) \big|_{t_1 = t_2 = t}
    \vt 
    \\
    \label{eq22}
    v_{2,t} &\equiv v_{2,t} (t,x_1,x_2) = v_2 ( \mtx_1, \mtx_2) \big|_{t_1 = t_2 = t }
    \vt 
\end{align}
This is the simplest choice for evaluating the trajectories, although we emphasise that we are not constrained by it. As discussed previously, from an operational standpoint it is a natural way of viewing an experiment performed in a particular reference frame with detectors tracking the evolution of each photon trajectory by performing successive detections on an ensemble. Moreover, we are additionally motivated by the possibility of a field-theoretic formulation of our measurement-based trajectories, in which the standard approach (i.e.\ in QFT) uses a single time coordinate \cite{mandel1995optical}. We also note that the multipartite velocity equations derived here satisfy a natural nonrelativistic limit, reducing to the equations one would obtain if beginning with Wiseman's nonrelativistic framework. The nonrelativistic limit is derived in the Methods. 

Equations (\ref{eq31}) and (\ref{eq32}) are manifestly Lorentz-covariant, each satisfying the velocity addition rule. That is, under a Lorentz boost to a reference frame with the constant velocity $\vartheta$, the components of $\mathbf{j}_i$ transform as 
\begin{align}
    j_i'(\mtx_1, \mtx_2) &= \gamma \left( j_i( \mtx_1, \mtx_2)  - \vartheta \rho_i(\mtx_1, \mtx_2) \right) ,
    \label{eq41}
    \vt 
    \\
    \rho_i' ( \mtx_1, \mtx_2) &= \gamma \left( \rho_i(\mtx_1, \mtx_2 ) - \vartheta j_i(\mtx_1, \mtx_2) \right) ,
    \label{eq42}
    \vt 
\end{align}
where $\gamma = 1/\sqrt{1-\vartheta^2}$ is the Lorentz factor, which yields 
\begin{align}\label{eq20}
    v_i' (\mtx_1, \mtx_2) &= \frac{v_i (\mtx_1, \mtx_2 ) - \vartheta }{1 - \vartheta v_i(\mtx_1, \mtx_2)} .
\end{align}
The velocity fields each transform according to the relativistic addition rule, in accordance with the Lorentz covariant structure of the multiparticle Klein-Gordon theory. Specifically, Eq.\ (\ref{eq20}) describes the velocity of particle $i$ in a new reference frame boosted by $\vartheta$, given it had the velocity $v_i$ in the original frame. Implicit in our notation is that the corresponding Lorentz transformations are applied to the coordinates $\mathbf{X}_{1,2}$ on the left-hand side of the equation. Moreover, the boosted expressions for $j_i'$ and $\rho_i'$ satisfy corresponding continuity equations in the boosted coordinates $(t_i',x_i')$. Of course, the equal time evolution parameter $t$ defined in Eq.\ (\ref{eq21}) and (\ref{eq21}) will no longer correspond to an equal timeslice in the new reference frame. We discuss the details of how the resulting trajectories are  related to each other in the section ``Boosts.''

\subsection{Equivalence of the Measurement-Based and Multiparticle Klein-Gordon Approaches}
We can now demonstrate the equivalence between the velocity fields derived using our measurement-based approach compared with those obtained via the multiparticle relativistic Klein-Gordon theory. First, 
note that in the measurement-based approach, the velocity fields can be simplified as follows:
\begin{align}
    v_1 (t,x_1,x_2) &= \frac{2 \mathrm{Re} \: \langle \psi(t) | \bar{x} \rangle \langle \bar{x} | \hat{\mathbf{k}}_A | \psi(t) \rangle}{2 \mathrm{Re} \: \langle \psi(t) | \bar{x} \rangle \langle \bar{x} | \hat{\mathbf{H}}_A | \psi(t) \rangle},
    \\
    v_2 (t,x_1,x_2) &= \frac{2 \mathrm{Re} \: \langle \psi(t) | \bar{x} \rangle\langle \bar{x} | \hat{\mathbf{k}}_B | \psi(t) \rangle}{2 \mathrm{Re} \: \langle \psi(t) | \bar{x} \rangle \langle \bar{x} | \hat{\mathbf{H}}_B | \psi(t) \rangle }.
\end{align}
In this form, it is straightforward to show that the multitime Klein-Gordon conserved currents and densities, evaluated on a single timeslice $t_1 = t_2 = t$, are equivalent to the respective terms in the weak value-based velocity fields. That is, 
\begin{align}
    j_1(\mtx_1,\mtx_2) \big|_{t_1 = t_2 = t} &= 2 \mathrm{Re} \: \langle \psi(t) | \bar{x} \rangle\langle \bar{x} | \hat{\mathbf{k}}_A | \psi(t) \rangle , 
    \vt 
    \\
    j_2( \mtx_1, \mtx_2) \big|_{t_1 = t_2 = t} &= 2\mathrm{Re} \: \langle \psi(t) | \bar{x} \rangle\langle \bar{x} | \hat{\mathbf{k}}_B | \psi(t) \rangle 
    , 
    \vt 
    \\
    \label{eq46}
    \rho_1(\mtx_1,\mtx_2) \big|_{t_1 = t_2 = t} &= 2 \mathrm{Re} \: \langle \psi(t) | \bar{x} \rangle \langle \bar{x} | \hat{\mathbf{H}}_A | \psi(t) \rangle 
    , 
    \vt 
    \\
    \label{eq47}
    \rho_2( \mtx_1, \mtx_2) \big|_{t_1 = t_2 = t} &= 2 \mathrm{Re} \: \langle \psi(t) | \bar{x} \rangle \langle \bar{x} | \hat{\mathbf{H}}_B | \psi(t) \rangle 
    \vt ,
\end{align}
implying
\begin{align}
    v_{i\mathrm{M}}(t,x_1,x_2) &= v_{i\mathrm{KG}}(\mtx_1,\mtx_2) \big|_{t_1 = t_2 = t} 
\end{align}
where $i = 1,2$ and we have specially denoted the velocity fields as derived from the measurement-based and Klein-Gordon approaches by $v_{i\mathrm{M}}$ and $v_{i\mathrm{KG}}$ respectively. This is the main result of this paper. That is, the relativistic Bohmian-type velocity fields for a position- and time-symmetrised Klein-Gordon wavefunction (this wavefunction satisfying two continuity equations associated with the spacetime variables of each photon) can be operationally grounded in weak measurements enacted by detectors that are agnostic to the identity of the individual photons. We also reiterate that though we have considered the simplest nontrivial interaction between two particles, our weak value framework is generally applicable to more complex interactions involving multiple particles. Now using this equivalence, it is possible for us to obtain the relativistic multiparticle trajectories of the photons.

\subsection{Physical Interpretation of the Klein-Gordon Densities}
To plot the trajectories, we first need to obtain analytic expressions for the conserved currents and densities of the respective photons, the explicit forms of which are derived in ``Methods.'' We then numerically solve the coupled differential equations $v_{i,t} = \D x_i/\D t$ given some initial position for each photon and plot the resulting curves. Similarly to the single-photon trajectories derived in \cite{Foo_2022Rel}, we consider two-photon trajectories in the so-called optical approximation, where the Gaussian wavepackets in the initial state have wavevectors centred far away from the origin. In this regime and if the particles are indistinguishable (i.e.\ equal variances and centre frequencies), then the densities obtained via the multiparticle Klein-Gordon dynamics can be shown to be equal:
\begin{align}\label{eq5.60}
    \rho_1(\mtx_1, \mtx_2) \big|_{t_1 = t_2 = t} &= \rho_2 (\mtx_1, \mtx_2 ) \big|_{t_1 = t_2 = t}.
\end{align}
More importantly, there exists an equivalence between the energy-averaged sum of these densities and that predicted by the measurement-based ``field-theoretic'' approach i.e.\ the probability density obtained from the standard two-particle amplitude $\left| \langle \bar{x} | \psi(t) \rangle \right|^2 \equiv \left| \psi_\mathrm{M}(t,x_1,x_2) \right|^2$ (the general proof of which is shown in Methods):
\begin{align}\label{eq5.61}
    \big| \psi_\mathrm{M}(t,x_1,x_2) \big|^2 &= \frac{\rho_1(\mtx_1, \mtx_2) + \rho_2(\mtx_1, \mtx_2) }{4k_0} \bigg|_{t_1 = t_2 = t}
\end{align}
for $k_0 \gg \sigma$. The form of the right-hand side is suggestive of the fact that the Klein-Gordon densities associated with each particle should be interpreted physically as \textit{energy densities}, rather than probability densities. The sum of these densities, normalised by the sum of the mean energies of the two particles, $4k_0$, gives the same result as the two-particle amplitude obtained via the standard field-theoretic approach. The latter is clearly a positive-definite quantity, allowing for the interpretation of Eq.\ (\ref{eq5.61}), in the optical approximation, as a genuine probability density. Moving beyond the optical approximation likely requires an extension of our theory to a description of Bohmian mechanics that can handle states with non-conserved particle number \cite{durrPhysRevLett.93.090402,Nikolic2004QFT,Detlefdurr_2003,Struyve_2011}.  

\subsection{Relativistic Two-Photon Trajectories}\label{sec5.2.4}

\begin{figure}[t]
    \centering
    \includegraphics[width=0.725\linewidth]{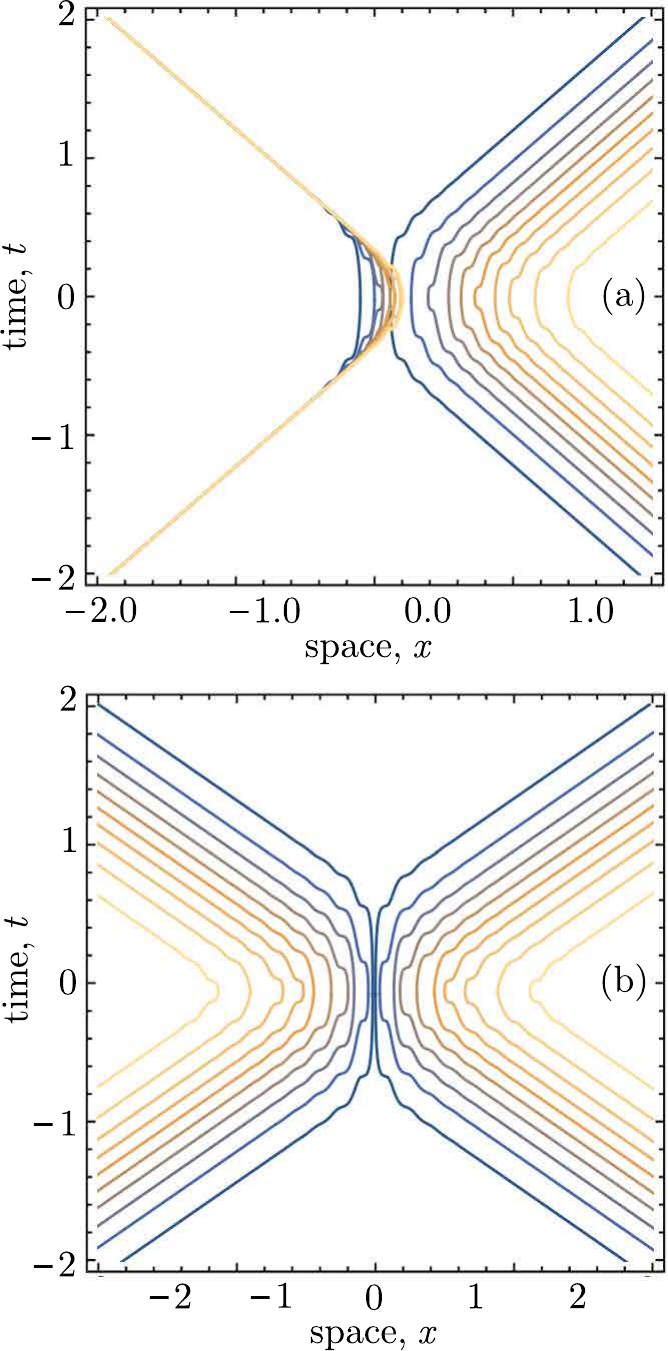}
    \caption[Two-photon Bohmian trajectories for a position- and time-symmetrised wavefunction]{Two-photon trajectories for (a) one of the photons at a fixed initial position, and (b) symmetric initial conditions. Each pair of trajectories in each colour corresponds to a given pair of initial conditions for both photons. To obtain these trajectories, we solved the set of differential equations $v_i (\mtx_1, \mtx_2) = \D x_i / \D t$ for a pair of initial positions $x_{i0} \equiv x_i(t_0)$. The time-evolution occurs on the equal time-slice $t_1 = t_2 = t$. We have utilised the parameters $k_0 / \sigma = 20$.}
    \label{fig:1}
\end{figure}

One instructive representation of the trajectories overlays the integral curves of the two velocity fields on a single spacetime diagram in the coordinates $(t,x)$, shown in Fig.\ \ref{fig:1}. Figure \ref{fig:1}(a) displays the photon trajectories when one of the photons is initially positioned at $x_1 = -2$, while the initial position of the other is Gaussian-distributed according to the position space representation of the probability density $\rho_{2,t}$. Figure \ref{fig:1}(b) plots pairs of trajectories with symmetric initial conditions. The trajectories display some interesting features. First, a given pair of trajectories never cross, meaning that the incident photons interact in such a way as to ``repel'' each other. Such behaviour, within the Bohmian interpretation of particles following classical trajectories, is entirely novel. Second, the behaviour of the photons upon near-approach suggests a nonclassical interference effect, which is further elucidated in Fig.\ \ref{fig:2bohm}. Even if one of the photons begins its trajectory at an identical initial position, the position of the second photon clearly has a nontrivial effect on its trajectory.

\begin{figure}[t]
\centering
    \includegraphics[width=0.725\linewidth]{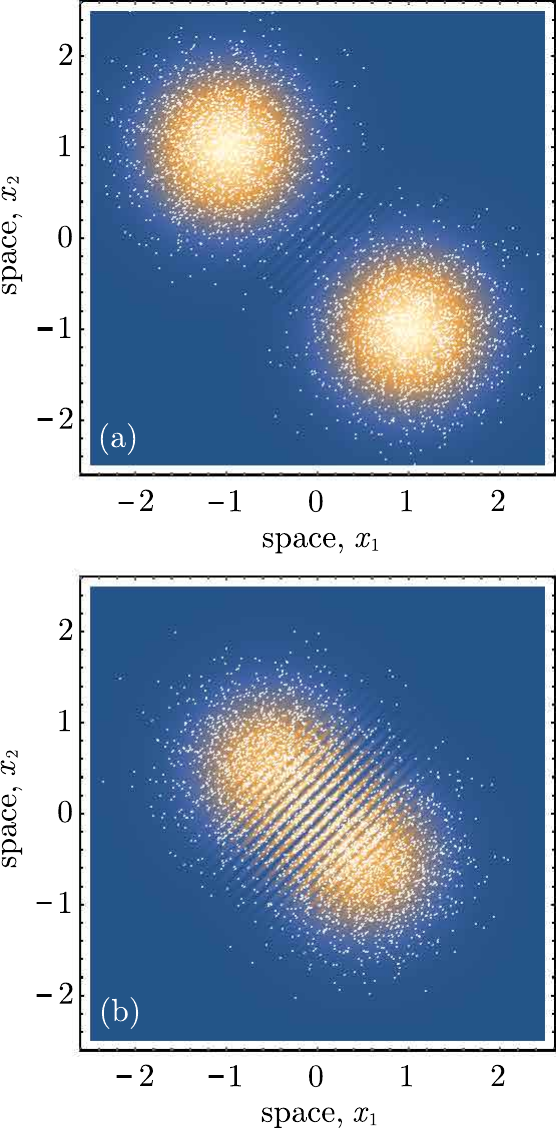}
    \caption[Individual timeslices of trajectories with randomly sampled initial conditions]{Individual timeslices of the trajectories plotted in the $(x_1,x_2)$ parameter space. The respective subplots display a random sample of trajectories that are initially distributed according to the probability density $\rho_t(t,x_1,x_2)$ at $t = -2$. The three timeslices correspond to (a) $t = -1$, (b) $t = -0.5$ and (c) $t = 0$. We have used $k_0/\sigma = 20$ as in Fig.\ \ref{fig:1}. }
    \label{fig:2bohm}
\end{figure}

A second representation of the trajectories is shown in Fig.\ \ref{fig:2bohm}. Here, we have plotted individual timeslices of a given pair of trajectories in the $(x_1,x_2)$ parameter space, each pair represented by a white dot. The symmetry of the density about the line $x_1 = x_2$ is a manifestation of the indistinguishability of the photons. We have randomly sampled a Gaussian distribution that reflects the initial distribution of the quantum-mechanical probability density, $\rho_t(t,x_1,x_2)$, in the $(x_1,x_2)$ parameter space. Moving into the interference region, we observe that the density of trajectories matches the probability density predicted by the multitime Klein-Gordon theory. The interference fringes that emerge are a manifestation of photon bunching, in which the position-correlated photons interfere with each other nonclassically \cite{bachor2019guide}. This effect arises because the detectors, when positioned within the interference region, cannot distinguish the identity of the particles. This is a purely quantum-mechanical interference effect. 

\subsection{Boosts}
To illustrate the Lorentz-covariant properties of the derived velocity fields and trajectories, we can consider these from a boosted reference frame. In the multiparticle Klein-Gordon theory, we can achieve this by mapping the individual spacetime points associated with a given trajectory using the standard Lorentz transformation:
\begin{align}
    (t,x) \mapsto \gamma( t - \vartheta x , x - \vartheta t ) .
\end{align}
Note in particular that one need not invoke the two position and time variables associated with the respective particles; each pair of spacetime coordinates $(t_1,x_1)$ or $(t_2,x_2)$ parametrising the coordinates of the individual particles can simply be considered some independent event occurring at $(t,x)$, to which we apply the above transformation. An identical result may be achieved by applying the relativistic velocity addition rule to the respective velocity fields Eq.\ (\ref{eq21}) and (\ref{eq22}), and utilising appropriately boosted initial conditions in the reference frame. The coupling of the two velocity fields means that the use of two position and time variables (in order to perform coordinate boosts of \textit{each particle}) is required to perform this transformation. Only \textit{after} the boost can one fix the new evolution parameter $t_1' = t_2' = t'$ to evolve along an equal timeslice (in the new frame), from which one can solve the resulting coupled differential equations in this new reference frame. Provided one correctly transforms the initial conditions from the original frame into the boosted one, then the trajectories obtained from either approach are identical. We give a numerical demonstration of this equivalence in the code provided at Ref./ \cite{code}.  

In the measurement-based approach, an effective reference frame transformation can be achieved by adjusting the wavevectors and bandwidths of the wavepackets by an appropriate redshift. That is, if in the original frame the wavevectors and bandwidths of the wavepackets were equal, $k_{0R} = k_{0L} = k_0$ and $\sigma_R = \sigma_L = \sigma$, then in the new frame these will receive an effective redshift given by 
\begin{align}
    \mathbf{k}_{0R} &= \mathbf{k}_0 \sqrt{\frac{1-\vartheta}{1+\vartheta}}, \:\:\:\:\:
    \mathbf{k}_{0L} = \mathbf{k}_0 \sqrt{\frac{1+\vartheta}{1-\vartheta}},
    \label{eq56}
\end{align}
where $\mathbf{k}_{0R} = (k_{0R},\sigma_R)$, $\mathbf{k}_{0L} = ( k_{0L}, \sigma_L)$, and $\mathbf{k}_0 = (k_0, \sigma)$. As discussed above, to ensure that the measurement-based and multiparticle Klein-Gordon approaches yield the same result (i.e.\ so that they describe the same physical setup), one needs to choose appropriate initial conditions for each pair of photons in the latter case, as obtained from the former. That is, a given pair of trajectories initialised at $x_1(t_0) = - x_2(t_0) = x_0$ on the timeslice $t = t_0$ will no longer be simultaneous in the boosted reference frame. To replicate the same physical scenario (from the measurement-based perspective), one needs to account for this when plotting pairs of trajectories. 

\begin{figure}[h]
    \centering
    \includegraphics[width=0.75\linewidth]{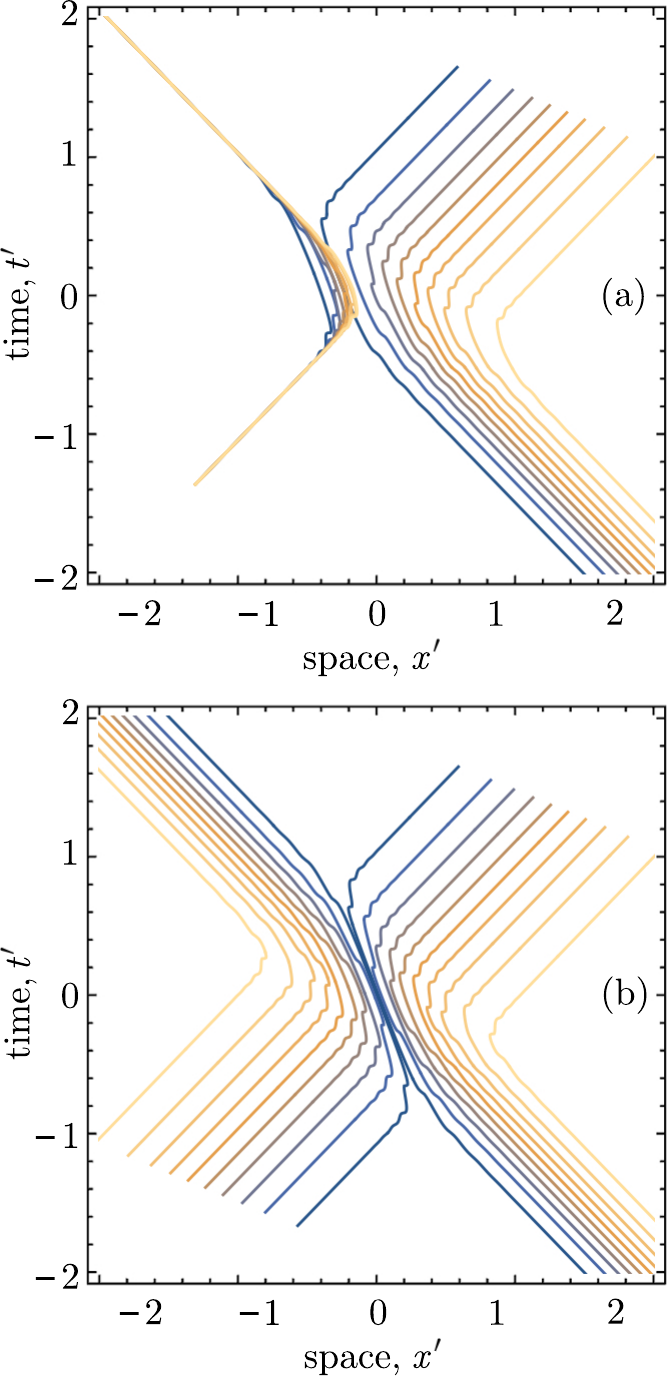}
    \caption[Two-photon Bohmian trajectories from a boosted reference frame]{Two-photon trajectories from a boosted reference frame with $\vartheta = 0.4$. Subplot (a) corresponds to the initial conditions shown in Fig.\ \ref{fig:1}(a), while (b) corresponds to the initial conditions shown in Fig.\ \ref{fig:1}(b). Both plots are obtained with a single evolution parameter $t'$. The different initial conditions for each trajectory reflects the relativity of simultaneity between the initial conditions in the original frame compared with those in the new frame.}
    \label{fig:3bohm}
\end{figure}

In Fig.\ \ref{fig:3bohm}, we have plotted the photon trajectories from a boosted reference frame with $\vartheta = 0.4$. The initial conditions of the particles are obtained from those used in the original reference frame from Figs.\ \ref{fig:1} and \ref{fig:2bohm}, so that they are no longer simultaneous in the new frame}. The path of the photons is now skewed with respect to the coordinates of the new reference frame. The same trajectories are obtained from both the measurement and the KG approaches.

\begin{figure}[h]
    \centering
    \includegraphics[width=0.75\linewidth]{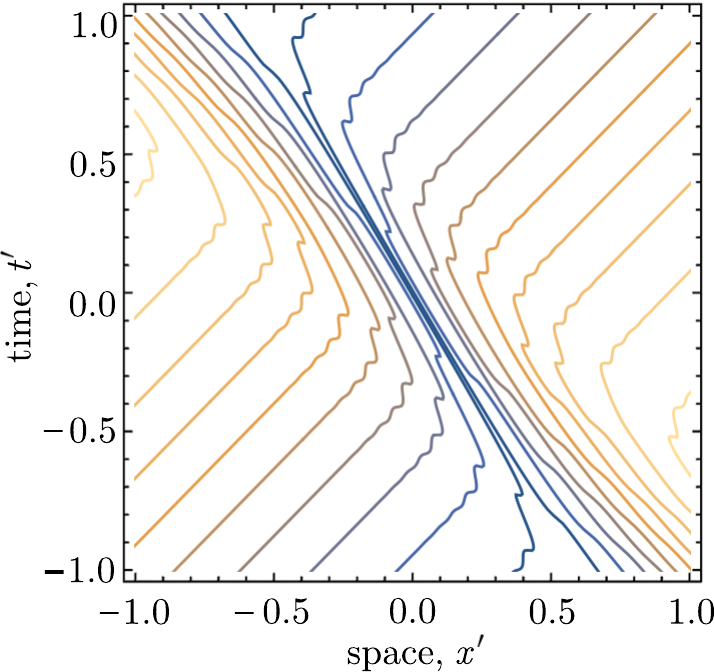}
    \caption{A zoomed-in view of the backwards-in-time trajectories for $\vartheta = 0.6$. }
    \label{fig:3bohmboost}
\end{figure}

For sufficiently large boosts, the trajectories can also move backwards-in-time according to the observer in this reference frame, see Fig.\ \ref{fig:3bohmboost}. As we noted in \cite{Foo_2022Rel}, this is a feature of our construction of the velocity field as a coordinate velocity obtained in a particular reference frame, and thus entirely consistent with the tenets of general relativity. As long as the optical approximation is satisfied, $k_{0R(L)} \gg \sigma_{R(L)}$, there always exists a reference frame in which the velocity field is positively directed everywhere.

The trajectories in the boosted reference frame also satisfy quantum-mechanical continuity. For two photons with equal energy in the original reference frame, and assuming the optical limit, the energy weak value associated with each detector (equivalently, the Klein-Gordon conserved current density associated with each particle) are equal, Eq.\ (\ref{eq5.60}) and (\ref{eq5.61}). In the boosted reference frame, this will not be true in general. Operationally, this is reflective of the fact that the photons with different energies are no longer indistinguishable according to the detectors. It is likewise consistent with our previous interpretation of the Klein-Gordon densities $\rho_{i,t'}$ as \textit{energy densities}. 

Now, we wish to confirm the consistency of the derived trajectories in the boosted reference frame with quantum-mechanical continuity. As before, we need to add the respective Klein-Gordon energy densities and normalise by the mean energy of the combined two-photon state. This is the same procedure performed previously to obtain the trajectories shown in Fig.\ \ref{fig:2bohm}, although in that case the photons were indistinguishable with respect to both their spacetime location and their energies. The energy-normalised density is the actual particle density predicted by the multiparticle Klein-Gordon theory, and is by construction symmetric with respect to an exchange of the particle labels. Crucially, this is also the result obtained via the standard two-particle amplitude obtained in QFT, $| \langle \bar{x} | \psi(t) \rangle |^2 \equiv | \psi_\mathrm{M}(t,x_1,x_2) |^2 $, upon applying the optical approximation. That is, in the optical approximation, the following equality holds:
\begin{align}\label{eq5.92}
    \big| \psi_\mathrm{M}(t,x_1,x_2) \big|^2 &= \frac{\rho_1(\mtx_1, \mtx_2) + \rho_2(\mtx_1, \mtx_2) }{2(k_{0R}+k_{0L})} \bigg|_{t_1 = t_2 =t} 
\end{align}
for $k_{0R,L} \gg \sigma_{R,L}$, in the general scenario when the centre frequencies and variances of the two photons are not equal, which we prove in Methods. Equation (\ref{eq5.92}) is the generalisation of Eq.\ (\ref{eq5.61}) for photons with different mean energies (a special case of which is physically equivalent to a Lorentz boost to a new reference frame). Thus, in the optical approximation, the energy-normalised average of the Klein-Gordon densities is identical to the standard probability density obtained using the measurement-based approach, even for Lorentz boosted scenarios. The consistency of the derived trajectories is thus confirmed by the satisfaction of continuity with respect to this probability density.

\begin{figure}[h]
    \centering
    \includegraphics[width=0.675\linewidth]{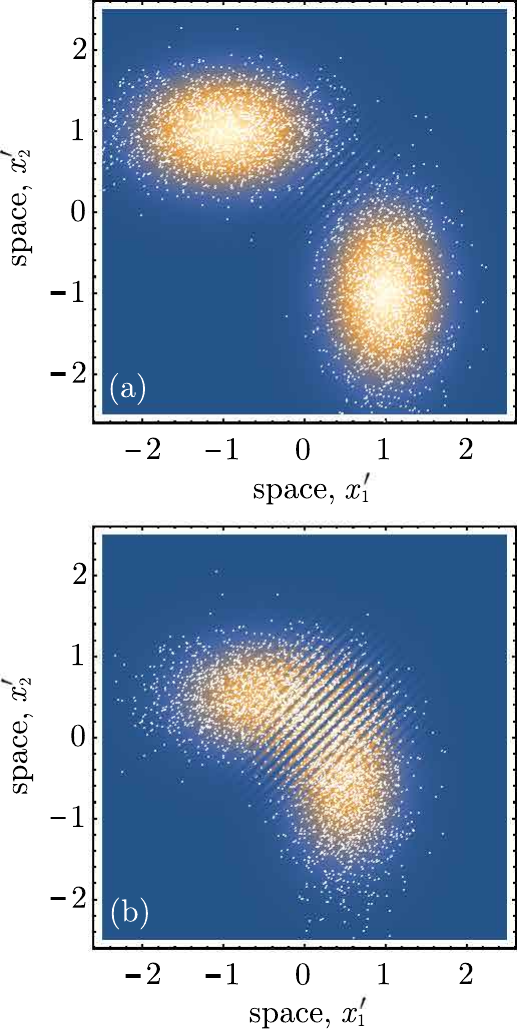}
    \caption[Individual timeslices of the trajectories from a boosted reference frame]{Individual timeslices of the trajectories plotted in the $(x_1',x_2')$ parameter space. The respective subplots display the boosted initial conditions utilised in Fig.\ \ref{fig:2bohm}, for a boost of $\vartheta = 0.2$. We have used the wavepacket parameters $k_0 /\sigma = 20$. }
    \label{fig:6}
\end{figure}

This consistency is shown graphically in Fig.\ \ref{fig:6}, where we have plotted two timeslices of the boosted trajectories in the $(x_1',x_2')$ parameter space. The density of trajectories matches the probability density predicted by the measurement-based approach, underlaid in each subplot. The boost induces redshifts in the wavepacket variances, broadening and contracting the width of the wavepackets outside the interference region.

\section{Discussion}
In this article, we have generalised a weak value framework for relativistic Bohmian trajectories of photons to interactions involving multiple photons. We have shown, contrary to widely held belief, that a trajectory-based ontology for relativistic particles is consistent with relativity when one bases the trajectories in detector measurements of momentum and energy, albeit with a nonlocal dependence on the particle position. In particular, we studied a nontrivial interaction in which the agnosticism of the detectors to the identity of the particles resulted in nonclassical correlations and hence interference between them. We connected the two-photon velocity fields, obtained via weak measurements, with those derived using a multitime scalar Klein-Gordon theory, in which the wavefunction is symmetrised in its spacetime location. This choice of the wavefunction was enforced \textit{a priori} by the fact that the two photons are indistinguishable, and ultimately yields an identical result to that obtained using our operationally based method. 

In \cite{Foo_2022Rel}, we gave a general relativistic interpretation of the photon trajectories in terms of a so-called ``quantum metric.'' As an extension to the quantum potential in nonrelativistic Bohmian mechanics (sourced by the wavefunction and enacting a force upon the particles that guides their trajectories), we demonstrated that a modified Alcubierre metric \cite{Alcubierre_1994} gives rise to the exact trajectories predicted by our weak value velocity for single particles, including those with spacelike tangents that become backwards-in-time according to Lorentz boosted observers. In particular, the metric giving rise to the line element, 
\begin{align}
    \D s^2 &= - (1 - v_s^2 ) \D t^2 - 2 v_s \D x \D t + \D x^2 
\end{align}
was shown, through an appropriate choice of $v_s$, to yield a coordinate velocity for photons that corresponded with that obtained via the weak-value formalism:
\begin{align}
    v(t,x) \equiv \frac{\D x}{\D t} = \frac{j(t,x)}{\rho(t,x)}.
\end{align}
This general idea of absorbing the effect of the quantum potential into the geometric structure of relativistic spacetime was originally conjectured by Pitowsky \cite{Pitowsky1991} and Shojai et.\ al.\ \cite{shojaidoi:10.1142/S0217751X98000305,jalalzadehdoi:10.1142/S0217732319502705}, though, these authors did not consider an explicit spacetime that could generate actual trajectories. 

Here, we can extend the proposal of Ref.\ \cite{Foo_2022Rel} to a metric based in local evaluations of the wavefunction and its derivatives, giving rise to an interpretation of the two-photon trajectories via a curved spacetime geometry. Let us consider the Alcubierre-like geometry with line element (metric), 
\begin{align}\label{eq60}
    \D s^2 &= - ( 1 - \mathbf{v}_s^2 ) \D t^2 - 2 \mathbf{v}_s \D x \D t + \D x^2 ,
\end{align}
where
\begin{align}
    \mathbf{v}_s &= \left( \left| \frac{j_i ( \psi(\mtx_1, \mtx_2) , \p^x_i \psi (\mtx_1, \mtx_2) ) }{\rho_i(\psi(\mtx_1, \mtx_2) , \p^t_i \psi (\mtx_1, \mtx_2))} \right| - 1 \right) 
    \non 
    \\
    & \qquad \times \mathrm{sgn} \left( \frac{j_i(\psi(\mtx_1, \mtx_2) , \p^x_i (\mtx_1, \mtx_2))}{\rho_i(\psi(\mtx_1, \mtx_2) , \p^t_i(\mtx_1, \mtx_2)) } \right) .
    \label{eq61}
\end{align}
In Eq.\ (\ref{eq61}), $\p^{x}_i$ and $\p^{t}_i$ denote derivatives evaluated locally at the spacetime point of the particle of interest ($i = 1 , 2$). From Eq.\ (\ref{eq60}), it can be straightforwardly shown that the coordinate velocity of particle $i$ propagating in this metric is 
\begin{align}
    v_i(\mtx_1, \mtx_2) &= \frac{2 \mathrm{Im} \: \psi^\star(\mtx_1, \mtx_2) \p^{x}_i \psi(\mtx_1, \mtx_2)}{2\mathrm{Im} \: \psi^\star(\mtx_1, \mtx_2) \p^{t}_i \psi(\mtx_1, \mtx_2)} , 
    \nonumber 
    \\
    &\equiv \frac{j_i(\mtx_1, \mtx_2)}{\rho_i(\mtx_1,\mtx_2)},
\end{align}
where $j_i(\mtx_1, \mtx_2)$ and $\rho_i(\mtx_1, \mtx_2)$ represent the conserved current and density of the Klein-Gordon wavefunction evaluated locally at the position $(t_i, x_i)$ of the particle of interest. In this sense, Eq.\ (\ref{eq60}) represents the local curvature at the spacetime point of either of the photons. From the derived equivalence between the measurement-based and Klein-Gordon representations of the velocity fields, this curvature can in-principle be inferred from local weak measurements of the energy and momentum. We note that the proposed metric has a nonlocal dependence on the spacetime location of both particles. It was pointed out in Ref.\ \cite{bravermanPhysRevLett.110.060406} that such a dependence, taken alone, does not require a nonlocal explanation of the trajectories. This is because Bohmian trajectories are deterministic, and such ``nonlocal'' dependence of the velocity (here guided by a relativistic metric) can be interpreted as an unusual dependence on the initial conditions of the photons. In Methods, we propose a physical setup involving a spacelike intervention that would demonstrate unequivocally the nonlocal character of the guiding metric. 

The nonlocality of the guiding metric is a curious feature of our results. In particular, we have shown that the derived photon trajectories--considered ``independently'' from any metric interpretation--obey Lorentz covariance, as one would expect for relativistic particles. Further, by grounding the velocity fields and resulting trajectories in a measurement protocol based in weak values, one anticipates such trajectories to be observable in experiment, notwithstanding the technical difficulties accompanying such an endeavour. It is in this sense that the proposed relativistic trajectories are ``consistent,'' and we expect that such consistency should hold in more complicated scenarios involving, for example, Bell-inequality-violating setups. The trajectories that arise in such setups, and the interpretation of the resulting guiding metrics, remains an open question left for future research.

\section{Methods}
\subsection{Derivation of Weak Value Velocity Expressions}
Let us first compute the wavefunction given that the strong measurement of position is symmetrised due to the agnosticism of the two detectors. We have (omitting a normalisation factor of $1/\sqrt{2}$ for brevity), 
\begin{widetext}
\begin{align}
    \langle \bar{x} | \psi(t) \rangle &= \left( \int\D k_{1_A} e^{ik_{1_A} x_1} \langle k_{1_A} | \int\D k_{2_B} e^{ik_{2_B}x_2} \langle k_{2_B} | \right) \left( \int\D k_1 e^{-i|k_1|t} f(k_1;k_{0R}) | k_1 \rangle \int\D k_2 e^{-i|k_2|t}f(k_2;k_{0L}) | k_2 \rangle \right)
    \non 
    \\
    & + \left( \int\D k_{2_A} e^{ik_{2_A}x_1} \langle k_{2_A} | \int\D k_{1_B} e^{ik_{1_B}x_2} \langle k_{1_B} | \right) \left( \int\D k_1 e^{-i|k_1|t}f(k_1;k_{0R}) | k_1 \rangle \int\D k_2 e^{i|k_2|t}f(k_2;k_{0L}) |k_2 \rangle \right) ,
    \non 
    \\
    \langle \bar{x} | \psi(t) \rangle &= \int\D k_{1_A} \int\D k_1 e^{-i|k_1|t+ik_{1_A}x_1} f(k_1;k_{0R}) \delta(k_{1_A} - k_1) \int\D k_{2_B} \int\D k_2 e^{-i|k_2|t + ik_{2_B} x_2} f(k_2;k_{0L}) \delta(k_{2_B} - k_2) 
    \non 
    \\
    & + \int\D k_{1_B} \int\D k_1 e^{-i|k_1|t+ ik_{1_B} x_2} f(k_1;k_{0R} ) \delta(k_{1_B} - k_1) \int\D k_{2_A} \int\D k_2 e^{-i|k_2|t + ik_{2_A} x_1} f(k_2;k_{0L} ) \delta(k_{2_A} - k_2 ) ,
    \non 
    \\
\label{eq5.37}
    \langle \bar{x} | \psi(t) \rangle &=  \int\D k_1 e^{-i|k_1|t+ik_{1}x_1} f(k_1;k_{0R}) \int\D k_2 e^{-i|k_2|t + ik_{2} x_2} f(k_2;k_{0L}) 
    \non 
    \\
    & \qquad + \int\D k_1 e^{-i|k_1|t+ ik_{1} x_2} f(k_1;k_{0R} ) \int\D k_2 e^{-i|k_2|t + ik_{2} x_1} f(k_2;k_{0L} ) ,
    \non 
\\
    \langle \bar{x} | \psi(t) \rangle &= \int\D k_1 e^{-ik_1(t - x_1)} f(k_1;k_{0R}) \int\D k_2 e^{-ik_2(t + x_2)} f(k_2;k_{0L}) 
    \non 
    \\
    & \qquad + \int\D k_1 e^{-ik_1(t - x_2)} f(k_1;k_{0R}) \int\D k_2 e^{-ik_2(t + x_1) } f(k_2;k_{0L}) .
\end{align}
\end{widetext}
where in the last line, we have applied the optical approximation, where the right- and left-moving wavepackets are only supported on the regions $k>0$ and $k<0$ respectively. Clearly, the wavefunction is symmetric under an exchange of the photon positions, $x_1 \Leftrightarrow x_2$. Let us now compute the numerators for each of the weak value expressions, Eq.\ (\ref{eq5.31})-(\ref{eq5.34}). Taking $\langle \bar{x} | \hat{\mathbf{H}}_A | \psi(t) \rangle$ for example, there are four contributing terms:
\begin{widetext}
\begin{align}
    T_1 &= \int\D k_{1A} e^{ik_{1A}x_1} \langle k_{1A} | \int\D k_{2B} e^{ik_{2B}x_2} \langle k_{2B} | \Big( \hat{H}_A | k_{1A} \rangle\langle k_{1A} | \otimes \hat{I}_B | k_{2B} \rangle\langle k_{2B} | \Big) | \psi(t) \rangle ,
    \\
    T_2 &= \int\D k_{1A} e^{ik_{1A}x_1} \langle k_{1A} | \int\D k_{2B} e^{ik_{2B}x_2} \langle k_{2B} | \Big( \hat{H}_A |k_{2A} \rangle\langle k_{2A} | \otimes \hat{I}_B |k_{1B} \rangle\langle k_{1B} | \Big) |\psi(t) \rangle ,
    \\
    T_3 &= \int\D k_{2A} e^{ik_{2A}x_1} \langle k_{2A} | \int\D k_{1B} e^{ik_{1B}x_2} \langle k_{1B} | \Big( \hat{H}_A | k_{1A} \rangle\langle k_{1A} | \otimes \hat{I}_B |k_{2B} \rangle\langle k_{2B} | \Big) | \psi(t) \rangle ,
    \\
    T_4 &= \int\D k_{2A} e^{ik_{2A} x_1} \langle k_{2A} | \int\D k_{1B} e^{ik_{1B}x_2} \langle k_{1B} | \Big( \hat{H}_A |k_{2A} \rangle\langle k_{2A} | \otimes\hat{I}_B | k_{1B} \rangle \langle k_{1B} | \Big) | \psi(t) \rangle .
\end{align}
Let us calculate these terms explicitly. First, we have, 
\begin{align}
    T_1 &=  \int\D k_{1A} e^{ik_{1A}x_1}\langle k_{1A}| \int\D k_{2B} e^{ik_{2B}x_2} \langle k_{2B} | \Big( \hat{H}_A | k_{1A} \rangle\langle k_{1A}  |\otimes \hat{I}_B |k_{2B} \rangle\langle k_{2B} |  \Big) , \non 
    \non 
    \\
    & \qquad  \int\D k_1' e^{-i|k_1'|t}f(k_1';k_{0R}) |k_1'\rangle \int\D k_2' e^{-i|k_2'|t }f(k_2';k_{0L}) |k_2' \rangle ,
    \\
    &= \int \D k_{1A} e^{ik_{1A}x_1} \langle k_{1A} | \hat{H}_A | k_{1A} \rangle \int\D k_{2B} e^{ik_{2B}x_2} \langle k_{2B} | \hat{I}_B | k_{2B} \rangle 
    \non 
    \\
    & \qquad \int\D k_{1}'e^{-i|k_1'|t}f(k_1';k_{0R}) \delta(k_{1A} - k_1') \int\D k_2 e^{-i|k_2'|t}f(k_2';k_{0L}) \delta( k_{2B} - k_2') .
\intertext{To obtain the second line, we have used the fact that the momentum eigenstates associated with the detectors $A$ and $B$ live in separate Hilbert spaces. Next, using the result $\langle k_{1A}| \hat{H}_A | k_{1A} \rangle = |k_{1A}|$, we have}
    T_1 &= \int\D k_{1A} e^{-i|k_{1A}|t + i k_{1A} x_1} |k_{1A} | f(k_{1A};k_{0R}) \int\D k_{2B} e^{-i|k_{2B}| t + ik_{2B}x_2} f(k_{2B};k_{0L}). 
\intertext{Next}
    T_2 &= \int\D k_{1A} e^{ik_{1A}x_1}\langle k_{1A}| \int\D k_{2B} e^{ik_{2B}x_2} \langle k_{2B} | \Big( \hat{H}_A | k_{2A} \rangle\langle k_{2A}  |\otimes \hat{I}_B |k_{1B} \rangle\langle k_{1B} | \Big) 
    \non 
    \\
    & \qquad  \int\D k_1' e^{-i|k_1'|t}f(k_1';k_{0R}) |k_1'\rangle \int\D k_2' e^{-i|k_2'|t }f(k_2';k_{0L}) |k_2' \rangle , 
    \non 
    \\
    &= \int\D k_{1A} e^{ik_{1A}x_1} \langle k_{1A} | \hat{H}_A | k_{2A} \rangle \int\D k_{2B} e^{ik_{2B}x_2} \langle k_{2B} | \hat{I}_B | k_{1B} \rangle
    \non 
    \\
    & \qquad \int\D k_1' e^{-i|k_1'|t}f(k_1';k_{0R}) \langle k_{1B} | k_1' \rangle \int\D k_2' e^{-i|k_2'|t}f(k_2';k_{0L})\langle k_{2A} | k_2' \rangle = 0 ,
\intertext{where in the last equality we have used the fact that $\langle k_{1A} | \hat{H}_A | k_{2A} \rangle = 0$ (that is, $\hat{H}_A$ is diagonal in the basis $\{ |k_{1A} \rangle , | k_{2A} \rangle \} $). Similarly,}
    T_3 &= \int\D k_{2A} e^{ik_{2A} x_1} \langle k_{2A} | \int\D k_{1B} e^{ik_{1B}x_2} \langle k_{1B} | \Big( \hat{H}_A | k_{1A} \rangle\langle k_{1A} | \otimes\hat{I}_B | k_{2B} \rangle\langle k_{2B} | \Big) 
    \non 
    \\ 
    & \qquad \int\D k_1 e^{-i|k_1'
    |t }f(k_1';k_{0R})|k_1' \rangle \int\D k_2' e^{-i|k_2'|t} f(k_2';k_{0L}) | k_2' \rangle ,
    \\
    &= \int\D k_{2A} e^{ik_{2A}x_1} \langle k_{2A} | \hat{H}_A | k_{1A} \rangle \int\D k_{1B} e^{ik_{1B}x_2} \langle k_{1B} | \hat{I}_B | k_{2B} \rangle
    \non 
    \\
    & \qquad \int\D k_1' e^{-i|k_1'|t} f(k_1';k_{0R}) \langle k_{1A} | k_1' \rangle \int\D k_2' e^{-i|k_2'|t}f(k_2';k_{0L}) \langle k_{2B} | k_2' \rangle = 0 ,
    \intertext{where as with $T_2$, we have used the fact that $\langle k_{2A} | \hat{H}_A | k_{1A} \rangle = 0$. Finally,}
    T_4 &= \int\D k_{2A} e^{ik_{2A} x_1} \langle k_{2A} | \int\D k_{1B} e^{ik_{1B}x_2} \langle k_{1B} | \Big( H_A | k_{2A} \rangle\langle k_{2A} | \otimes \hat{I}_B | k_{1B} \rangle\langle k_{1B} | \Big) 
    \non 
    \\
    & \qquad \int\D k_1' e^{-i|k_1'|t} f(k_1';k_{0R}) | k_1' \rangle \int\D k_2' e^{-i|k_2'|t} f(k_2';k_{0L}) |k_2 '\rangle ,
    \non 
    \\
    &= \int\D k_{2A} e^{ik_{2A}x_1} \langle k_{2A} | \hat{H}_A | k_{2A} \rangle \int\D k_{1B} e^{ik_{1B}x_2} \langle k_{1B} | \hat{I}_B | k_{1B} \rangle \non 
    \\
    & \qquad \int\D k_1' e^{-i|k_1'|t}f(k_1';k_{0R}) \langle k_{1B} | k_1' \rangle \int\D k_2' e^{-i|k_2'|t}f(k_2';k_{0L}) \langle k_{2A} | k_2' \rangle ,
    \non 
    \\
    &= \int\D k_{2A} e^{-i|k_{1A}|t + ik_{1B}x_2} f(k_{1B};k_{0R}) \int\D k_{2A} e^{-i|k_{2A} |t + ik_{2A} x_1} |k_{2A} | f(k_{2A};k_{0L}).
\end{align}
The total expression is thus, 
\begin{align}
    \langle \bar{x} | \hat{\mathbf{H}}_A | \psi(t) \rangle &= \int\D k\: e^{-i|k|t + ikx_1}|k|f(k;k_{0R}) \int\D k' \: e^{-i|k'|t + ik'x_2}f(k';k_{0L}) 
    \non 
    \\
    & \qquad + \int\D k \: e^{-i|k|t + ikx_2}f(k;k_{0R}) \int\D k' \: e^{-i|k'|t + ik'x_1}|k'|f(k';k_{0L}). 
\intertext{In the optical approximation, this simplifies to}
    \langle \bar{x} | \hat{\mathbf{H}}_A | \psi(t) \rangle &= \int\D k \: e^{-ik(t - x_1)} kf(k;k_{0R}) \int\D k' e^{-ik'(t + x_2)} f(-k';k_{0L}) 
    \non 
    \\
    & \qquad + \int\D k \: e^{-ik(t - x_2)} f(k;k_{0R}) \int\D k' e^{-ik'(t + x_1)} k' f(-k';k_{0L}) .
\end{align}
By applying the same reasoning to the other three weak value expressions, we obtain the result stated in the main text. 
\end{widetext}

\subsection{Derivation of the Klein-Gordon Velocity Expressions}
To obtain the Klein-Gordon wavefunction, we first define the right- and left-moving single particle states
\begin{align}
    | \psi_1 (t) \rangle &= \int \D k \: e^{-i|k|t} f_1(k;k_{0R}) | k \rangle ,
    \\
    | \psi_2 (t) \rangle &= \int\D k \: e^{-i|k|t} f_2(k;k_{0L}) | k \rangle ,
\end{align}
where the wavepackets have strong support on $k>0$ and $k<0$ respectively:
\begin{align}
    f_1(k;k_{0R}) &= \mathcal{N}_R \exp \left[ - \frac{(k-k_{0R})^2}{4\sigma_R^2} \right],
    \\
    f_2(k;k_{0L}) &= \mathcal{N}_L \exp \left[ - \frac{(k + k_{0L})^2}{4\sigma_L^2} \right].
\end{align}
In the position basis,
\begin{align}
    \langle x | \psi_1(t) \rangle &= \int\D k \: e^{-i|k|t + i k x} f_1(k;k_{0R}) ,
    \\
    \langle x | \psi_2(t) \rangle &= \int\D k \: e^{-i|k|t + i k x} f_2(k;k_{0L})  .
\end{align}
The position- and time-symmetrised wavefunction is 
\begin{align}
    \psi_\mathrm{KG} \equiv \psi_\mathrm{KG}(\mtx_1, \mtx_2) &= \frac{1}{\sqrt{2}} \left( \psi_1(\mtx_1) \psi_2(\mtx_2) + \psi_1(\mtx_2) \psi_2(\mtx_1) \right).
\end{align}
Explicitly (again dropping the $1/\sqrt{2}$ for brevity), we have:
\begin{widetext}
\begin{align}
    \psi_\mathrm{KG}(\mtx_1, \mtx_2) &= \int\D k_1 \: e^{-i|k_1|t_1 + i k_1 x_1}f_1(k_1;k_{0R})   \int\D k_2 \: e^{-i|k_2|t_2 + ik_2 x_2} f_2(k_2;k_{0L})   \non 
    \\
    & \qquad + \int\D k_1 \: e^{-i|k_1|t_2 + ik_1 x_2} f_1(k_1;k_{0R})  \int\D k_2 \: e^{-i|k_2|t_1 + ik_2x_1}f_2(k_2;k_{0L})  , \non 
    \\ 
    &= \int\D k_1 e^{-ik_1(t_1 - x_1)} f_1(k_1;k_{0R}) \int\D k_2 e^{-ik_2(t_2 + x_2)} f_2(-k_2;k_{0L}) 
    \non 
    \\
    & \qquad + \int\D k_1 e^{-ik_1(t_2 - x_2)} f_1(k_1;k_{0R}) \int\D k_2 e^{-ik_2(t_1 + x_1)} f_2(k_2;k_{0L}) . 
\end{align}
\end{widetext}
Evaluating this on the equal timeslice $t_1 = t_2 = t$ gives the same result as that obtained using the operational approach. The time and space derivatives of the wavefunction can straightforwardly be shown to be 
\begin{widetext}
\begin{align}
    \frac{\p \psi_\mathrm{KG}}{\p t_1} &= - i \int\D k_1 \: e^{-ik_1(t_1 - x_1)} f_1 (k_1;k_{0R}) k_1  \int\D k_2 \: e^{-ik_2(t_2 + x_2)} f_2(-k_2;k_{0L})  
    \non 
    \\
    & \qquad - i \int\D k_1 \: e^{-ik_1(t_2 - x_2)} f_1(k_1;k_{0R}) \int\D k_2 \: e^{-ik_2(t_1 + x_1)}f_2(-k_2;k_{0L}) k_2 ,
\\
    \frac{\p \psi_\mathrm{KG}}{\p t_2} &= - i \int\D k_1 \: e^{-ik_1(t_1 - x_1)} f_1 (k_1;k_{0R}) \int\D k_2 \: e^{-ik_2(t_2 + x_2)} f_2(-k_2;k_{0L})  k_2 
    \non 
    \\
    & \qquad  - i \int\D k_1 \: e^{-ik_1(t_2 - x_2)} f_1(k_1;k_{0R}) k_1 \int\D k_2 \: e^{-ik_2(t_1 + x_1)}f_2(-k_2;k_{0L}) ,
\\  
    \frac{\p \psi_\mathrm{KG}}{\p x_1} &= i \int\D k_1 e^{-ik_1(t_1 - x_1)} f_1(k_1;k_{0R}) k_1 \int\D k_2 e^{-ik_2(t_2 + x_2)} f_2(-k_2;k_{0L})
    \non
    \\
    & \qquad  - i \int\D k_1 e^{-ik_1(t_2 - x_2)} f_1(k_1;k_{0R}) \int\D k_2 e^{-ik_2(t_1 + x_1)} f_2(-k_2;k_{0L}) k_2 ,
\\
    \frac{\p \psi_\mathrm{KG}}{\p x_2} &= - i \int\D k_1 e^{-ik_1(t_1 - x_1)} f_1(k_1;k_{0R}) \int\D k_2 e^{-ik_2(t_2 + x_2)} f_2(k_2;k_{0L}) k_2 
    \non 
    \\
    & \qquad + i \int\D k_1 e^{-ik_1(t_2 - x_2)} f_1(k_1;k_{0R}) k_1 \int\D k_2 e^{-ik_2(t_1 + x_1)} f_2(-k_2;k_{0L}) .
\end{align}
\end{widetext}
These derivatives are used to compute the time and space components of the conserved current vector for the respective photons.

\subsection{Proof of Eq.\ (\ref{eq5.92})}\label{sec:proof}
We wish to demonstrate that the energy-normalised Klein-Gordon densities yield the same prediction as that obtained via a ``field-theoretic'' approach in the optical approximation. That is, in the optical approximation, we want to prove that
\begin{align}\label{eq5.93}
    \big| \psi_\mathrm{KG}(\mtx_1, \mtx_2 ) \big|^2 &= \frac{\rho_1(\mtx_1, \mtx_2 ) + \rho_2(\mtx_1,\mtx_2)}{2(k_{0R} + k_{0L})} 
\end{align}
for $k_{0R,L} \gg \sigma_{R,L}$, in the general scenario where the photons have unequal centre frequencies and variances, where
\begin{align}
    \rho_i &= 2 \mathrm{Im} \: \psi^\star(\mtx_1, \mtx_2 ) \p_i^t \psi(\mtx_1, \mtx_2)
\end{align}
are the usual time-components of the conserved current vectors associated with particles $i = 1 , 2$ and $\psi_\mathrm{KG}(\mtx_1 , \mtx_2)$ is the position- and time-symmetrised wavefunction, 
\begin{align}
    & \psi_\mathrm{KG}(\mtx_1, \mtx_2) =
    \non 
    \vt 
    \\
    & \:\:\: \frac{1}{\sqrt{2}} \left( \psi_1(\mtx_1) \psi_2( \mtx_2) + \psi_1 ( \mtx_2) \psi_2(\mtx_1) \right). 
\end{align}
Evaluating Eq.\ (\ref{eq5.93}) on an equal timeslice $t_1 = t_2 = t$ yields the condition stated in the main text, Eq.\ (\ref{eq5.92}):
\begin{align}
    \big| \psi_\mathrm{M}(t,x_1,x_2) \big|^2 &= \frac{\rho_1(\mtx_1, \mtx_2) + \rho_2(\mtx_1, \mtx_2) }{2(k_{0R}+k_{0L})} \bigg|_{t_1 = t_2 = t} 
\end{align}
for $k_{0R,L} \gg \sigma_{R,L}$.
To begin the proof, we associate, as usual, the labels $\psi_1$ and $\psi_2$ with the Klein-Gordon wavefunctions describing a right- and left-moving Gaussian wavepacket respectively:
\begin{align}
\label{eqA.3}
    \psi_1(t,x) &= \int\D k \: e^{-ik(t - x)} f_R(k) ,
    \\
\label{eqA.4}
    \psi_2(t,x) &= \int\D k \: e^{-ik(t + x)} f_L(k) ,
\end{align}
where as usual, our wavepackets take the form
\begin{align}
    f_R(k) &= \mathcal{N}_R \exp \Big[ - \frac{(k-k_{0R})^2}{4\sigma_R^2} \Big] ,
    \\
    f_L(k) &= \mathcal{N}_L \exp \Big[ - \frac{(k-k_{0L})^2}{4\sigma_L^2} \Big] . 
\end{align}
Performing the integrals of Eq.\ (\ref{eqA.3}) and (\ref{eqA.4}), we obtain, 
\begin{widetext}
\begin{align}
    \psi_1(t,x) &= \left( \frac{2\sigma_R^2}{\pi} \right)^{1/4} \exp \Big[ - (t- x) \left( i k_{0R} + (t - x ) \sigma_R^2 \right) \Big] ,
    \\
    \psi_2(t,x) &= \left( \frac{2\sigma_L^2}{\pi} \right)^{1/4} \exp \Big[ - (t + x) \left( i k_{0L} + (t + x) \sigma_L^2 \right) \Big] .
\end{align}
The analytic expression for the position- and time-symmetrised wavefunction is thus,
\begin{align}
    \psi_\mathrm{KG}(\mtx_1, \mtx_2) &= \sqrt{\frac{\sigma_R\sigma_L}{\pi}} \Big( \exp \left[ - (t_2 + x_2) \left( i k_{0L} + (t_2 + x_2) \sigma_L^2 \right) - (t_1 - x_1 ) \left( i k_{0R} + (t_1 - x_1) \sigma_R^2 \right) \right] 
    \non 
    \\
    & \qquad + \exp \left[ - (t_1 + x_1 ) \left( i k_{0L} + (t_1 + x_1) \sigma_L^2 \right) - (t_2 - x_2) \left( i k_{0R} + (t_2 - x_2) \sigma_R^2 \right) \right] \Big). 
\end{align}
The modulus square of this wavefunction gives the left-hand side of Eq.\ (\ref{eq5.93}), specifically:
\begin{align}
    \big| \psi_\mathrm{KG} (\mtx_1, \mtx_2) \big|^2 
    \vphantom{\sqrt{\frac{2}{\pi}}} &\simeq \frac{\sigma_R \sigma_L}{\pi} \Big( \exp \Big[ - 2 \left( V_2^2 \sigma_L^2 + U_1^2 \sigma_R^2 \right) \Big] + \exp \Big[ - 2 \left( V_1^2 \sigma_L^2 + U_2^2 \sigma_R^2 \right) \Big] 
     \vphantom{\sqrt{\frac{2}{\pi}}}
    \non 
    \\
    & \qquad + 2 \exp \Big[ - \left( V_1^2 + V_2^2 \right) \sigma_L^2 - \left( U_1^2 + U_2^2 \right) \sigma_R^2 \Big] \cos \Big[ k_{0R}(U_1 - U_2 ) - k_{0L}( V_1 - V_2) \Big] \Big)  , \vphantom{\sqrt{\frac{2}{\pi}}}
\end{align}
\end{widetext}
having defined $V_i = t_i + x_i$ and $U_i = t_i - x_i$ as lightcone coordinates. On the other hand, we find that the two Klein-Gordon (energy) densities are given by
\begin{widetext}
\begin{align}
    \rho_1(\mtx_1,\mtx_2) \non \vphantom{\sqrt{\frac{2}{\pi}}}
    &= \frac{2\sigma_R \sigma_L}{\pi} \big( \exp \left[ - 2\left( V_1^2 \sigma_L^2 + U_2^2 \sigma_R^2 \right) \right] k_{0L} + \exp \left[ - 2( V_2^2 \sigma_L^2 + U_1^2 \sigma_R^2) \right] k_{0R} 
    \non \vphantom{\sqrt{\frac{2}{\pi}}}
    \\
    &  \qquad   + \exp \left[ - \left( V_1^2 + V_2^2 \right) \sigma_L^2 - \left( U_1^2 + U_2^2 \right) \sigma_R^2 \right] ( k_{0L} + k_{0R} ) \cos \big[ k_{0R}(U_1 - U_2 ) - k_{0L}(V_1 - V_2) \big] \vphantom{\sqrt{\frac{2}{\pi}}}
    \non 
    \\
    &  \qquad   + 2 \exp \left[ - \left( V_1^2 + V_2^2 \right) \sigma_L^2 - \left( U_1^2 + U_2^2 \right) \sigma_R^2 \right] (V_1 \sigma_L^2 - U_1 \sigma_R^2) \sin \big[ k_{0R} (U_1 - U_2 ) - k_{0L}(V_1 - V_2) \big] \vphantom{\sqrt{\frac{2}{\pi}}} ,
    \label{eq101}
    \\
    \rho_2(\mtx_1,\mtx_2) \non \vphantom{\sqrt{\frac{2}{\pi}}}
    &= \frac{2\sigma_R \sigma_L}{\pi} \big( \exp \left[ - 2\left( V_2^2 \sigma_L^2 + U_1^2 \sigma_R^2 \right) \right] k_{0L} + \exp \left[ - 2( V_1^2 \sigma_L^2 + U_2^2 \sigma_R^2 \right] k_{0R} \vphantom{\sqrt{\frac{2}{\pi}}}
    \non 
    \\
    &  \qquad  + \exp \left[ - \left( V_1^2 + V_2^2 \right) \sigma_L^2 - \left( U_1^2 + U_2^2 \right) \sigma_R^2 \right] (k_{0L} + k_{0R})  \cos \big[ k_{0R}(U_1 - U_2 ) - k_{0L}(V_1 - V_2) \big] \vphantom{\sqrt{\frac{2}{\pi}}}
    \non 
    \\
    &  \qquad  - 2 \exp \left[ - \left( V_1^2 + V_2^2 \right) \sigma_L^2 - \left( U_1^2 + U_2^2 \right) \sigma_R^2 \right] ( V_2 \sigma_L^2 - U_2 \sigma_R^2 ) \sin \big[ k_{0R} (U_1 - U_2 ) - k_{0L}(V_1 - V_2) \big] \vphantom{\sqrt{\frac{2}{\pi}}} .
\end{align}
\end{widetext}
For completeness, the Klein-Gordon conserved currents associated with the two photons are given by, 
\begin{widetext}
\begin{align}
    j_1(\mtx_1, \mtx_2) \non \vphantom{\sqrt{\frac{2}{\pi}}}
    &= \frac{2\sigma_R \sigma_L}{\pi} \big( \exp \left[ - 2 \left( V_2^2 \sigma_L^2 + U_1^2 \sigma_R^2 \right) \right] k_{0R} - \exp \left[ -2 \left( V_1^2 \sigma_L^2 + U_2^2 \sigma_R^2 \right) \right] k_{0L} \vphantom{\sqrt{\frac{2}{\pi}}}
    \non 
    \\
    & \qquad + \exp \left[ - \left( V_1^2 + V_2^2 \right) \sigma_L^2 - \left( U_1^2 +  U_2^2 \right) \sigma_R^2 \right] \left( k_{0R} - k_{0L} \right) \cos \big[ k_{0L}(V_1 - V_2 ) - k_{0R} ( U_1 - U_2 ) \big] \vphantom{\sqrt{\frac{2}{\pi}}}
    \non 
    \\
    & \qquad + 2 \exp \left[ - \left( V_1^2 + V_2^2 \right) \sigma_L^2 - \left( U_1^2 + U_2^2 \right) \sigma_R^2 \right] \left( V_1 \sigma_L^2 + U_1 \sigma_R^2 \right) \sin \big[ k_{0L} (V_1 - V_2 ) - k_{0R} ( U_1 - U_2 ) \big] \big) \vphantom{\sqrt{\frac{2}{\pi}}} ,
    \label{eq102}
    \\
    j_2(\mtx_1, \mtx_2 ) \non \vphantom{\sqrt{\frac{2}{\pi}}}
    &= \frac{2\sigma_R \sigma_L}{\pi} \big( \exp \left[ - 2 \left( V_1^2 \sigma_L^2 + U_2^2 \sigma_R^2 \right) \right] k_{0R} - \exp \left[ - 2 \left( V_2^2 \sigma_L^2 + U_1^2 \sigma_R^2 \right) \right] k_{0L} 
    \non 
    \vphantom{\sqrt{\frac{2}{\pi}}}
    \\
    & \qquad + \exp \left[ \left( V_1^2 + V_2^2 \right) \sigma_L^2 - \left( U_1^2 + U_2^2 \right) \sigma_R^2 \right] (k_{0R} - k_{0L} ) \cos \big[ k_{0L}(V_1 - V_2 ) - k_{0R} (U_1 - U_2 ) \big] 
    \non 
    \vphantom{\sqrt{\frac{2}{\pi}}}
    \\
    & \qquad - 2 \exp \left[ - \left( V_1^2 + V_2^2 \right) \sigma_L^2 - \left( U_1^2 + U_2^2 \right) \sigma_R^2 \right] \left( V_2 \sigma_L^2 + U_2 \sigma_R^2 \right) \sin \big[ k_{0L} ( V_1 - V_2 ) - k_{0R} ( U_1 - U_2 ) \big]
    \vphantom{\sqrt{\frac{2}{\pi}}}.
\end{align}
Returning to the expressions for the Klein-Gordon densities, their sum is given by  
\begin{align}
    \rho_1(\mtx_1, \mtx_2) + \rho_2 ( \mtx_1, \mtx_2) \non 
    \vphantom{\sqrt{\frac{2}{\pi}}}
    &= \frac{2\sigma_R \sigma_L}{\pi} (k_{0R} + k_{0L}) \big( \exp \left[ - 2 \left( V_2^2 \sigma_L^2 + U_1^2 \sigma_R^2 \right) \right] + \exp \left[ - 2 \left( V_1^2 \sigma_L^2 + U_2^2 \sigma_R^2 \right) \right] 
    \vphantom{\sqrt{\frac{2}{\pi}}}
    \non 
    \\
    & \qquad + 2 \exp \left[ - \left( V_1^2 + V_2^2 \right) \sigma_L^2 - \left( U_1^2 + U_2^2 \right) \sigma_R^2 \right] \cos \big[ k_{0R}(U_1 - U_2 ) - k_{0L}( V_1 - V_2) \big] 
    \non 
    \\
    & \qquad + 2 \exp \big[ - \left(V_1^2 + V_2^2 \right) \sigma_L^2 - \left( U_1^2 + U_2^2 \right) \sigma_R^2 \big] \frac{ (V_1-V_2) \sigma_L^2 - (U_1-U_2) \sigma_R^2}{k_{0R} + k_{0L}} 
    \vphantom{\sqrt{\frac{2}{\pi}}}
    \non 
    \\
    & \qquad \qquad \times \sin \big[ k_{0R}(U_1 - U_2 ) - k_{0L}(V_1 - V_2 ) \big] \big) 
    \vphantom{\sqrt{\frac{2}{\pi}}}.
\end{align}
In the optical approximation, then the coefficient of the $\sin$ term is small, leaving us with:
\begin{align}
    \rho_1(\mtx_1, \mtx_2) + \rho_2(\mtx_1, \mtx_2) \non 
    \vphantom{\sqrt{\frac{2}{\pi}}}
    &\simeq \frac{2\sigma_R \sigma_L}{\pi} (k_{0R} + k_{0L}) \big( \exp \left[ - 2 \left( V_2^2 \sigma_L^2 + U_1^2 \sigma_R^2 \right) \right] + \exp \left[ - 2 \left( V_1^2 \sigma_L^2 + U_2^2 \sigma_R^2 \right) \right] 
    \non \vphantom{\sqrt{\frac{2}{\pi}}}
    \\
    & \qquad + 2 \exp \left[ - \left( V_1^2 + V_2^2 \right) \sigma_L^2 - \left( U_1^2 + U_2^2 \right) \sigma_R^2 \right] \cos \big[ k_{0R}(U_1 - U_2 ) - k_{0L}( V_1 - V_2) \big] \big) \vphantom{\sqrt{\frac{2}{\pi}}}
\end{align}
Normalising by the relativistic normalisation factor $2(k_{0R}+ k_{0L})$ gives,
\begin{align}\label{eqa.15}
    \frac{\rho_1(\mtx_1, \mtx_2) + \rho_2(\mtx_1, \mtx_2) }{2(k_{0R}+k_{0L})} 
    \vphantom{\sqrt{\frac{2}{\pi}}}
    &\simeq \frac{\sigma_R \sigma_L}{\pi} \Big( \exp \Big[ - 2 \left( V_2^2 \sigma_L^2 + U_1^2 \sigma_R^2 \right) \Big] + \exp \Big[ - 2 \left( V_1^2 \sigma_L^2 + U_2^2 \sigma_R^2 \right) \Big] 
     \vphantom{\sqrt{\frac{2}{\pi}}}
    \non 
    \\
    & \qquad + 2 \exp \Big[ - \left( V_1^2 + V_2^2 \right) \sigma_L^2 - \left( U_1^2 + U_2^2 \right) \sigma_R^2 \Big] \cos \Big[ k_{0R}(U_1 - U_2 ) - k_{0L}( V_1 - V_2) \Big] \Big)  , \vphantom{\sqrt{\frac{2}{\pi}}}
    \\
    &=  \left| \psi_\mathrm{KG}(\mtx_1, \mtx_2 ) \right|^2 .
     \vphantom{\sqrt{\frac{2}{\pi}}} 
\end{align}
\end{widetext}
Upon evaluating Eq.\ (\ref{eqa.15}) along an equal timeslice $t_1 = t_2 = t$, we find that this is also equal to the result obtained via our measurement-based approach, $\left| \langle \bar{x} | \psi(t,x_1,x_2)\rangle \right|^2 \equiv \left| \psi_\mathrm{M}(t,x_1,x_2) \right|^2$, as desired. 

\subsection{Boosts of the Klein-Gordon Density}

Here, we show the relativistic covariance of the Klein-Gordon density, i.e.\ Eq.\ (\ref{eq42}). Let us consider these components for particle $1$, for brevity. Taking Eq.\ (\ref{eq101}) and (\ref{eq102}) and composing the quantity $\gamma( \rho_1(\mathbf{X}_1,\mathbf{X}_2) - v j_1(\mathbf{X}_1, \mathbf{X}_2) )$ gives
\begin{widetext} 
\begin{align}
    & \gamma ( \rho_1(\mathbf{X}_1, \mathbf{X}_2 ) - v j_1(\mathbf{X}_1 , \mathbf{X}_2 ) ) 
    \nonvt 
    \\
    & \:\:\: = \frac{2\sigma_R\sigma_L}{\pi} \bigg( \sqrt{\frac{1-v}{1+v}} \exp \left[ -2 \left(V_2^2 \sigma_L^2 + U_1^2 \sigma_R^2 \right) \right] k_{0R} + \sqrt{\frac{1+v}{1-v}} \exp \left[ -2 \left( V_1^2 \sigma_L^2 + U_2^2 \sigma_R^2 \right) \right] k_{0L} 
    \nonvt \\
    & \:\:\: + \exp \left[ - \left( V_1^2 + V_2^2 \right) \sigma_L^2 - \left( U_1^2 + U_2^2 \right) \sigma_R^2 \right] \Big( \sqrt{\frac{1+v}{1-v}} k_{0L} + \sqrt{\frac{1- v}{1 + v}} k_{0R} \Big) \cos \left[ k_{0R}(U_1 - U_2 ) - k_{0L} ( V_1 - V_2 ) \right] 
    \nonvt \\
    & \:\:\: -  2 \exp \left[ - \left(V_1^2 + V_2^2 \right) \sigma_L^2 - \left(U_1^2 + U_2^2 \right) \sigma_R^2 \right] \Big(  \sqrt{\frac{1 - v}{1 + v}} V_1 \sigma_L^2 - \sqrt{\frac{1+v}{1-v}} U_1 \sigma_R^2 \Big) \sin \left[ k_{0R} ( U_1 - U_2 ) - k_{0L}( V_1 - V_2 ) \right] \bigg) .
\end{align} 
Boosting the coordinates via the standard Lorentz transformation, yielding 
\begin{align} 
    U_i &= \sqrt{\frac{1-v}{1+v}} U_i',
    \nonvt 
    \\
    V_i &= \sqrt{\frac{1+v}{1-v}} V_i'
    \nonvt 
\end{align} 
and using the definition of the redshifted frequencies and variances in Eq.\ (\ref{eq56}), we find 
\begin{align}
    &\gamma(\rho_1(\mathbf{X}_1, \mathbf{X}_2) - v j_1 (\mathbf{X}_1, \mathbf{X}_2 ) ) 
    \nonvt 
    \\
    & \:\:\: = \frac{2\sigma_R'\sigma_L'}{\pi} \bigg( \exp \left[ -2 ( V_2'^2 \sigma_L'^2 + U_1'^2 \sigma_R'^2 ) \right] k_{0R}' + \exp \left[- 2 ( V_1'^2 \sigma_L'^2 + U_2'^2 \sigma_R'^2 ) \right] k_{0L}' \nonvt 
    \\
    & \:\:\: + \exp \left[ - \left( V_1'^2 + V_2'^2 \right) \sigma_L'^2 - ( U_1'^2 + U_2'^2 \big) \sigma_R'^2 \right] \left( k_{0L}' + k_{0R}' \right) \cos \left[ k_{0R}'\left(U_1' - U_2'\right) - k_{0L}' \left( V_1' - V_2' \right) \right] 
    \nonvt 
    \\
    & \:\:\: - 2 \exp \left[ - (V_1'^2 + V_2'^2 ) - (U_1'^2 + U_2'^2 ) \sigma_R'^2 \right] \left( V_1' \sigma_L'^2 - U_1' \sigma_R'^2 \right) \sin \left[ k_{0R}' (U_1' - U_2' ) - k_{0L}' (V_1' - V_2' ) \right] \bigg) 
    \equiv \rho_1'(\mathbf{X}_1, \mathbf{X}_2)  
\end{align}
as desired. This is simply the statement that the form of the conserved current density is identical in the boosted reference frame, with an effective redshift of the frequency and variance of the wavepacket with respect to the original frame i.e.\ relativistic covariance. An identical property of the conserved current can be similarly shown following the same steps. 
\\\\
\end{widetext}

\subsection{Proposal to Demonstrate the Nonlocality of the Photon Trajectories}
\begin{figure}[h]
    \centering
    \includegraphics[width=0.8\linewidth]{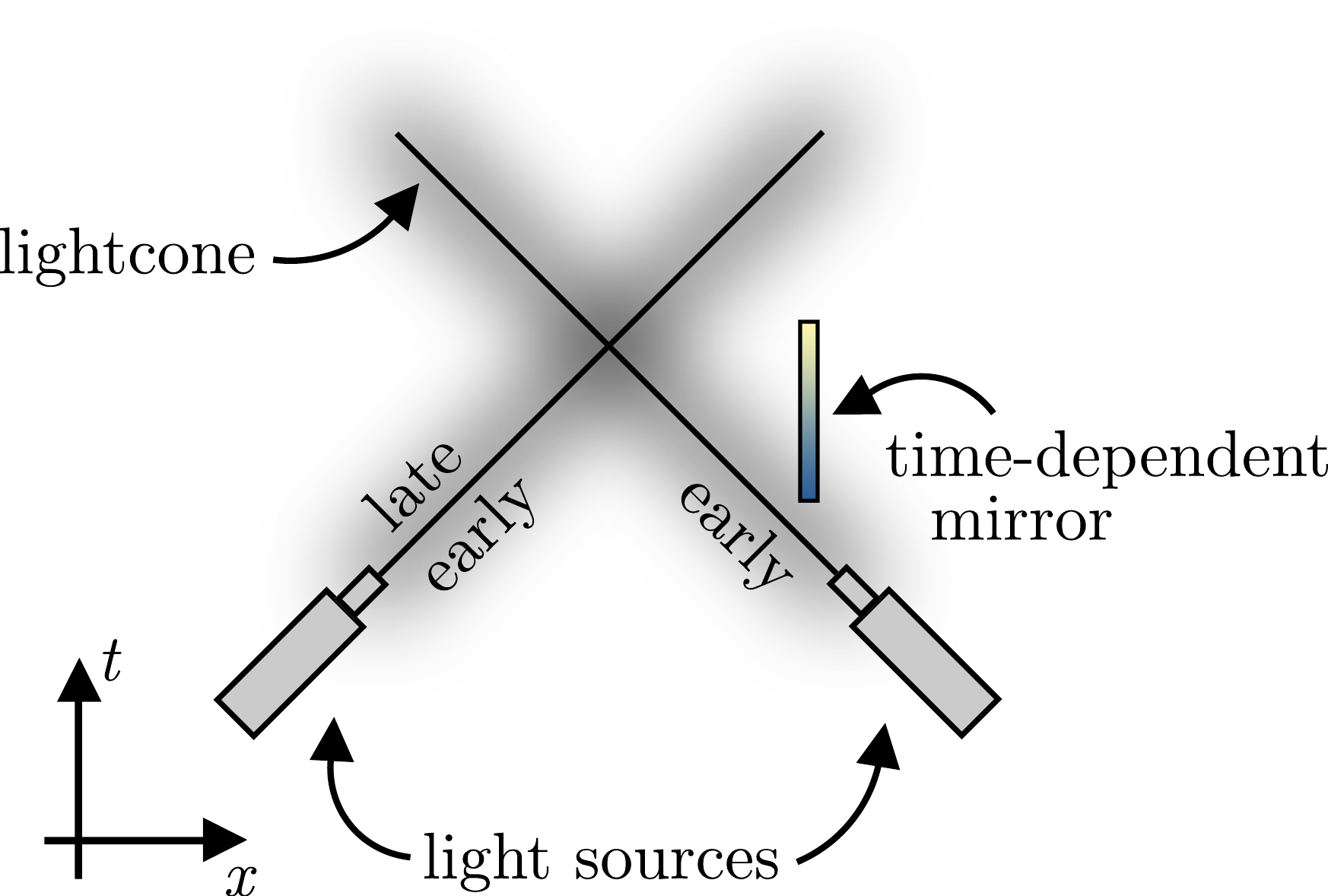}
    \caption{Schematic diagram of the physical setup that would demonstrate the nonlocality of the Bohmian trajectories, and hence the guiding metric.}
    \label{fig:7}
\end{figure}

In this section, we propose a physical setup that would demonstrate the nonlocality of the guiding metric. As mentioned in the Discussion, one needs to introduce a spacelike intervention on one of the photons and compute the resulting effect on the other photon. Such an intervention could take the form of a reflective mirror that is rapidly introduced at a fixed position, but only such that it affects the trailing edge of the initially left-moving wavepacket. This is depicted schematically in Fig.\ \ref{fig:7}.

As in our original setup, a pair of single-photon light sources are directed towards each other. Implied in Fig.\ \ref{fig:7} is the presence of detectors measuring the weak values of each photon in the manner prescribed in the Results. The main difference between the original system and that presented in Fig.\ \ref{fig:7} is the presence of an intervention, here depicted as a time-dependent mirror, enacted upon the trailing edge of the wavepacket for the left-moving mode. It is assumed that the mirror is rapidly introduced in such a way as to only affect the trailing edge of the wavepacket. In such a scenario, a pair of photons initially positioned in the leading edge of the wavepacket (one left-moving and one right-moving) will be unaffected by the mirror, and will follow the predetermined trajectories shown in blue in Fig.\ \ref{fig:1}. On the other hand, for an initially left-moving photon positioned in the trailing edge of the wavepacket (e.g.\ the light yellow trajectories in Fig.\ \ref{fig:1}), its trajectory will be redirected due to reflection off the mirror. Likewise, the trajectory of its partner (initially moving to the right) will be nontrivially affected due its dependence on the spacetime coordinates of the now-redirected photon. So as not to detract from the main results of the present work, we leave a detailed investigation of this scenario, which is naturally computationally expensive, for future work.

\subsection{Nonrelativistic Limit of the Velocity Equations}
In a similar manner to Ref.\ \cite{Foo_2022Rel}, our velocity equations for the two photons reduces to the standard nonrelativistic limit upon taking the so-called paraxial approximation, $k_z \gg k$. That is, $E(k) = \sqrt{k^2 + k_z^2} \simeq k_z + k^2/2k_z$ in the limit $k_z \gg k$. The single-particle wavefunctions in this regime are given by, 
\begin{align}
    \psi_1(t,x) &= \int\D k \: e^{-i \left( k_z + \frac{k^2}{2k_z} \right) t} e^{ikx} f_1(k;k_{0R}) ,  
    \\
    \psi_2(t,x) &= \int\D k \: e^{-i \left( k_z + \frac{k^2}{2k_z} \right)t} e^{ikx} f_2(k;k_{0L}) ,
\end{align}
which, in the position-symmetrised state of both photons, yields, 
\begin{align}
    \psi(\mtx_1, \mtx_2) &= \frac{1}{\sqrt{2}} ( \psi_1(\mtx_1) \psi_2(\mtx_2) + \psi_1(\mtx_2) \psi_2(\mtx_1)  ). 
\end{align}
The Klein-Gordon densities, as above, take the form, 
\begin{align}
    \rho_1(\mtx_1, \mtx_2) &= \mrm{Im} \bigg[ \psi_1^\star(\mtx_1 ) \psi_2^\star(\mtx_2) \frac{- \p }{\p t_1} \psi_1(\mtx_1 ) \psi_2(\mtx_2) 
    \nonvt 
    \\
    & + \psi_1^\star (\mtx_1 ) \psi_2^\star(\mtx_2) \frac{- \p }{\p t_1} \psi_1(\mtx_2) \psi_2(\mtx_1 ) 
    \nonvt 
    \\
    & + \psi_1^\star(\mtx_2) \psi_2^\star(\mtx_1 ) \frac{- \p }{\p t_1} \psi_1(\mtx_1) \psi_2(\mtx_2 ) 
    \nonvt 
    \\
    & + \psi_1^\star(\mtx_2) \psi_2^\star(\mtx_1) \frac{- \p}{\p t_1} \psi_1(\mtx_2)\psi_2(\mtx_1) \bigg] 
    \vt , 
    \\
    \rho_2(\mtx_1, \mtx_2) &= \mrm{Im} \bigg[ \psi_1^\star(\mtx_1 ) \psi_2^\star(\mtx_2) \frac{- \p }{\p t_2} \psi_1(\mtx_1 ) \psi_2(\mtx_2) 
    \nonvt 
    \\
    & + \psi_1^\star (\mtx_1 ) \psi_2^\star(\mtx_2) \frac{- \p }{\p t_2} \psi_1(\mtx_2) \psi_2(\mtx_1 ) 
    \nonvt 
    \\
    & + \psi_1^\star(\mtx_2) \psi_2^\star(\mtx_1 ) \frac{- \p }{\p t_2} \psi_1(\mtx_2) \psi_2(\mtx_2 ) 
    \nonvt 
    \\
    & + \psi_1^\star(\mtx_2) \psi_2^\star(\mtx_1) \frac{- \p}{\p t_2} \psi_1(\mtx_2)\psi_2(\mtx_1) \bigg] .
\end{align}
Let us look at the terms individually. The first term in $\rho_1$ takes the form,
\begin{widetext} 
\begin{align}
    T_1 &= \int\D k_1 \: e^{i \left( k_z + \frac{k_1^2}{2k_z} \right) t_1 - i k_1x_1 } f_1(k_1;k_{0R} ) \int\D k_2 \: e^{i\left( k_z + \frac{k_2^2}{2k_z} \right)t_2 - i k_2x_2} f_2(k_2;k_{0L} ) 
    \nonvt 
    \\
    & \qquad \int\D k_3 \: \left( i k_z + \frac{i k_3^2}{2k_z} \right) e^{-i\left( k_z + \frac{k_3^2}{2k_z} \right)t_1 + ik_3x_1} f_1(k_3;k_{0R} ) \int\D k_4 \: e^{-i\left( k_z + \frac{k_4^2}{2k_z} \right) t_2 + ik_4x_2} f_2(k_4;k_{0l} ) 
    \nonvt 
    \\
    &= \int\D k_1 \: e^{i \left( \frac{k_1^2}{2k_z} \right) t_1 - i k_1x_1 } f_1(k_1;k_{0R} ) \int\D k_2 \: e^{i\left( \frac{k_2^2}{2k_z} \right)t_2 - i k_2x_2} f_2(k_2;k_{0L} ) 
    \nonvt 
    \\
    & \qquad \int\D k_3 \: \left(  ik_z + \frac{i k_3^2}{2k_z} \right) e^{-i\left( \frac{k_3^2}{2k_z} \right)t_1 + ik_3x_1} f_1(k_3;k_{0R} ) \int\D k_4 \: e^{-i\left( \frac{k_4^2}{2k_z} \right) t_2 + ik_4x_2} f_2(k_4;k_{0l} ) 
    \nonvt 
    \\
    &= \int\D k_1 \D k_3 \: \left( i k_z + \frac{ik_3^2}{2k_z} \right) \exp \left[ \frac{i( k_1^2 - k_3^2) t_1 }{2k_z} - i (k_1 - k_3 ) x_1 \right]  f_1(k_1;k_{0R} ) f_1(k_3;k_{0R})
    \nonvt 
    \\
    & \qquad \int\D k_2 \D k_4 \: \exp \left[ \frac{i(k_2^2 - k_4^2)t_2}{2k_z} - i(k_2 - k_4)x_2 \right] f_2(k_2;k_{0L} ) f_2(k_4;k_{0L} ) 
    \nonvt 
    \\
    &= \int\D k_1 \D k_3 \:  \left( i k_z - \frac{i}{2k_z} \frac{\p^2}{\p x_1^2} \right) \exp \left[ \frac{i( k_1^2 - k_3^2) t_1 }{2k_z} - i (k_1 - k_3 ) x_1 \right] f_1(k_1;k_{0R} ) f_1(k_3;k_{0R})
    \nonvt 
    \\
    & \qquad \int\D k_2 \D k_4 \: \exp \left[ \frac{i(k_2^2 - k_4^2)t_2}{2k_z} - i(k_2 - k_4)x_2 \right] f_2(k_2;k_{0L} ) f_2(k_4;k_{0L} ) 
    \nonvt 
    \\
    &= i k_z \Big[ | \psi_1(\mtx_1 ) |^2 | \psi_2 (\mtx_2 ) |^2 \Big] - \frac{i}{2k_z} \psi_1^\star(\mtx_1) \frac{\p^2}{\p x_1^2} \psi_1(\mtx_1) | \psi_2(\mtx_2)|^2 
    \vt   
    \label{eq114}
\end{align}
\end{widetext} 
In the limit where the double spatial derivative of $\psi$ is negligible to $k_z$, one can neglect the second term in Eq.\ (\ref{eq114}), giving, 
\begin{align}
    T_1 &= i k_z \Big[ | \psi_1(\mtx_1) |^2 | \psi_2(\mtx_2) |^2 \Big] .
\end{align}
Similarly, for the other terms, one finds, 
\begin{align}
    T_2 &\simeq i k_z \psi_1^\star (\mtx_1) \psi_1(\mtx_2)  \psi_2^\star (\mtx_2 ) \psi_2(\mtx_1 ) , 
    \vt \\
    T_3 &\simeq i k_z \psi_1^\star(\mtx_2) \psi_1(\mtx_1) \psi_2^\star(\mtx_1) \psi_2(\mtx_2) , 
    \vt \\
    T_4 &\simeq i k_z | \psi_1 (\mtx_2) |^2 | \psi_2(\mtx_1) |^2 
    \vt , 
\end{align}
which means that $\rho_1$ (and, as can easily be verified, $\rho_2$) can be written in terms of the standard Schr\"odinger density (upon taking an equal time $t_1 = t_2 = t$), 
\begin{align}
    \rho_1(\mtx_1,\mtx_2) &= k_z | \psi(t,x_1,x_2) |^2 ,
    \vt 
\end{align}
in the paraxial (nonrelativistic) limit. The velocity equation of the two particles thus reduces to, 
\begin{align}
    v_i (\mtx_1 , \mtx_2) &= \frac{1}{k_z} \frac{\mrm{Im} \: \psi^\star (t,x_1,x_2) \p^x_i \psi(t,x_1,x_2)}{| \psi(t,x_1,x_2)|^2} ,
\end{align}
which corresponds with that obtained by Wiseman \cite{Wiseman_2007} in the nonrelativistic regime.

\section{Author Contributions}
J.F., A.P.L., and T.C.R.\ contributed to all aspects of the research.

\section{Competing Interests}
The authors declare no competing interests.

\bibliography{main}

\end{document}